\documentclass[a4paper,aps,prd,showpacs,twocolumn,nofootinbib] {revtex4}
\usepackage[T1]{fontenc}
\usepackage[utf8]{inputenc}
\usepackage[brazil,english]{babel}
\usepackage{tensor}
\usepackage[usenames,dvipsnames]{color}
  \definecolor{darkblue}{RGB}{0,0,150}
\usepackage{amsmath,amssymb}
\usepackage{tensor,relsize}
\usepackage{graphicx}
\usepackage{nicefrac}
\usepackage[unicode]{hyperref}
\usepackage{color}

\usepackage{multirow}

\usepackage{enumerate}

\newcommand{\ud}{\ensuremath{\mathrm{d}}}

\DeclareMathAlphabet{\mathpzc}{T1}{pzc}{m}{it}

\graphicspath{ {figuras/} }

\begin{document}

\title{Evolving black holes from conformal transformations of static solutions}
\author{Marina M. C. Mello, Alan  Maciel, Vilson T. Zanchin}

\affiliation{Centro de Ci\^{e}ncias Naturais e Humanas, 
Universidade Federal do ABC\\
Avenida dos Estados 5001, 09210-580 -- Santo Andr\'{e}, S\~{a}o Paulo, 
Brazil}

\begin{abstract}
A class of nonstationary spacetimes is obtained by means of a conformal 
transformation of the Schwarzschild metric, where the conformal factor $a(t)$ 
is an arbitrary function of the time coordinate only. We investigate several 
situations including some where the final state is a central object with 
constant mass. The metric is such that there is an initial big-bang type 
singularity and the final state depends on the chosen conformal factor. The 
Misner-Sharp mass is computed and a localized central object may be identified. 
The trapping horizons, geodesic and causal structure of the resulting 
spacetimes are investigated in detail. When $a(t)$ asymptotes to a constant in 
a short enough time scale, the spacetime presents an event horizon and its 
analytical extension reveals black-hole or white-hole regions. On the other hand, 
when $a(t)$ is unbounded from above as in cosmological models, the spacetime 
presents no event horizons and may present null singularities in the future.
The energy-momentum content and other properties of the respective spacetimes 
are also investigated. 

\end{abstract}


\maketitle

\section{Introduction}

The exact solutions of general relativity (GR) mostly used to model
realistic objects in astrophysics and in cosmology may roughly be divided into 
two classes: static/stationary vacuum solutions that model the gravitational 
field outside massive objects or black holes, and homogeneous expanding 
solutions that are used as models for the large scale structure of the
Universe. The first class aims to describe localized objects while the second 
one describes the global scale of the Universe. The fact that both 
class of solutions are successful in so different scales highlights that those 
scales are physically disconnected from each other. The interest in joining in 
one single picture these two aspects of gravitational physics is 
longstanding, 
as is the question of how bound gravitational systems are insensitive to the 
cosmological expansion in large scale. These issues can be traced to 
the McVittie proposal of a metric describing a point mass in an expanding 
universe in 1933 \cite{McVittie:1933zz}, the Einstein-Strauss model 
\cite{Einstein:1945id} and the Lemaitre-Tolman-Bondi models 
\cite{Tolman:1934za, bondi-1947}. These works founded the three main approaches 
to this issue, respectively, the crafting of analytical solutions with the 
wished asymptotics, the matching between the two types of solution, and the 
analysis of Einstein dynamical equations for some specified fluid content. 

The matching approach has led to the Swiss-cheese models, that allow to 
describe an expanding inhomogeneous universe filled with static bubbles 
that behave as if they were shielded from each other. Those models are 
interesting in order to study physics in an inhomogeneous universe 
\cite{Biswas:2007gi,Marra:2007pm} but they are too rigid to be considered the 
solution for the linking of local and global scales, because they rely strongly 
on the isotropy of the expansion and the spherical symmetry of the matching 
\cite{mars-2001,mena-2002}.

The dynamical analysis approach is the one that allows for the most realistic 
description of the transient regime between local and global scales, since a 
realistic fluid can be chosen as the source of the geometry, and the dynamical 
equations give the complete evolution of all physical quantities. However, the 
complexity of Einstein field equations is a formidable obstacle for the 
achievement of a general understanding of the transition between the global 
scale and the local scale behavior. With the simplifying assumption of 
spherical symmetry, some advance had been made, namely, the definition of a 
dynamically motivated separating surface between the two regimes, named matter 
trapping shells in the series of papers 
\cite{Mimoso:2009wj,ledelliou-2011,Delliou:2013xra,Mimoso:2013iga, 
Maciel:2015vva}.

In the present paper we follow the McVittie lineage. Namely, we study the 
properties of a class of solutions proposed by Thakurta \cite{Thakurta1981} as 
a model of cosmological black-hole spacetime, which are obtained by 
multiplying the Kerr solution by an expanding scale factor $a(t)$. However, the 
experience with the McVittie solution shows that identifying the content of a 
line element is not a simple task and the understanding of the causal structure 
of McVittie spacetimes had to wait about 80 years to be achieved since its 
publication \cite{Nolan:98,NolanII99,Nolan:1999wf,
Kaloper:2010ec,Lake:2011ni,daSilva:2012nh,daSilva:2015mja}.
Moreover, the McVittie metric 
was found to be richer than the original intent, with the possibility of 
representing black holes and/or white holes immersed in an expanding universe.
A possible explanation for such a lapse of time may be the little 
interest in the McVittie metric, but it may also be attributed to the lack of 
mathematical tools available for studying dynamical spacetimes. However, in 
the last two decades there was a great development in this field (see, e.g., 
\cite{Hayward:1993mw,Hayward:1993ph,Hayward:1994bu,Hayward:1997jp,Mars:2003ud, 
Senovilla:2011fk} and references there in), which has allowed for a 
systematical approach in order to understand dynamical solutions of GR, 
including cosmological black holes.

In this paper we make use of these tools in order to study thoroughly the 
nonrotating Thakurta solution, identifying all its possible outcomes with 
the 
main objective of finding examples of cosmological black holes. As the McVittie 
solution, the Thakurta solution can show a wide range of different properties, 
depending on the choice of the scale factor $a(t)$. For increasing and 
unbounded $a(t)$, as in cosmological models, we show that the Thakurta 
spacetime does not describe a cosmological black hole but rather an 
inhomogeneous expanding universe that may be either future geodesically 
complete or may present a future null singularity depending on the behavior of 
$a(t)$ for large times. On the other hand, for bounded $a(t)$, we establish 
sufficient conditions on $a(t)$ under which the Thakurta metric does present an 
event horizon at $r = 2m$. Moreover, we show that such an event horizon can be 
a black-hole horizon or a white-hole horizon depending on a further 
condition 
on $a(t)$.

This paper is organized as follows. In Sec.~\ref{sec:thakurta} we review the 
general properties of the Thakurta metric, analyze its source and the 
Misner-Sharp mass. In Sec.~\ref{sec:structural} we study the loci of 
coordinate and curvature singularities, the behavior of the trapping horizons 
and its dependency on the choice of $a(t)$, future and past geodesic 
completeness. Section~\ref{sec:r=2m} contains the properties of the locus 
given by $r = 2m$ and study the conditions on $a(t)$ that implies that this 
surface can either be a future null singularity or a traversable horizon, using 
this result to build cosmological black-hole models. In 
Sec.~\ref{sec:causal} we analyze thoroughly the possible types of conformal 
diagrams corresponding to different choices of $a(t)$, including the possible 
analytical extensions when the $r=2m$ surface is traversable, and discuss 
the physical content of each kind of solution. In
Sec.~\ref{sec:conclusion} we make further comments and conclude. 

Throughout the paper, derivatives with respect to the 
$t$ coordinate are denoted with an overhead dot. We use signature $(-,+,+,+)$ 
and natural units with $G=1=c$.

\section{The Thakurta metric}\label{sec:thakurta}

\subsection{Overview}

The proposal of this paper is to study the nonrotating class of
solutions emerging from the
metric presented by Thakurta in 1981 \cite{Thakurta1981}. The
metric originally was built as a conformal transformation of Kerr rotating
black-hole solution, whose conformal factor depends only on the 
Boyer-Lindquist
time coordinate. 
The Thakurta metric is given by
\begin{align}\label{eq:thakurta0}
\ud s^{2} & = a^2(\eta)\left[-\frac{\Delta}{\Sigma^2}\left(\ud \eta -
j\sin^2\theta\,\ud \phi\right)^2  + 
 \frac{ \Sigma^2\ud r^2}{\Delta} + \Sigma^2 d\theta^2\right. \nonumber\\
& \left.+ \frac{\sin^2\theta}{\Sigma^2}\Big[\left(r^2 + j^2\right)
\ud \phi - j\sin^2\theta\,\ud \eta\Big]^2\right],
\end{align}
%%%%%%%%%%%%
with $\Sigma=r^2+j^2\cos^2\theta$ and $\Delta=r^2+j^2-2mr$,
where $m$ and $j$ are constant parameters. In the case $a(\eta)=$  constant
metric Eq.~\eqref{eq:thakurta0} reduces to the Kerr metric, with $m$ and $j$
being respectively the mass and the angular momentum per unit mass of the
Kerr black-hole metric.
On the other hand, for large $r$ the Thakurta metric asymptotes to the
flat Friedmann-Lemaitre-Robertson-Walker (FLRW) cosmological metric.
Hence, the coordinate $\eta$ and the function $a(\eta)$ are 
respectively the conformal time and the scale factor of the asymptotic
FLRW metric.

Our proposal is to analyze  the nonrotating Thakurta metric, that is, we set 
the angular momentum $j$ to zero, so that the metric assumes the form
\begin{equation}  
\begin{split}
\ud s^2& =a^2(\eta)\left[-\left(1-\frac{2m}{r}
\right)\ud \eta^2+\frac{\ud r^2}{1-\dfrac{2m}{r}}+r^2 \ud
\Omega^2\right]\\
&=-\left(1-\dfrac{2m}{r}\right)\ud t^2+\frac{a^2(t)\,\ud r^2}{1-\dfrac{2m}{r}}
+a^2(t)\, r^2\ud\Omega^2, 
\end{split}
\label{eq:thakurta}
\end{equation}
where the cosmological time $t$, defined by  $\ud t=a(\eta)\,
\ud \eta$,  was introduced.

It is worth to write here the Thakurta
metric in terms of the areal radius coordinate $R = a(t)\, r$,
\begin{eqnarray} 
\ud s^2& =-\left(1-\dfrac{2M(t)}{R}-\dfrac{H^2(t)R^2}{1-\dfrac{2M(t)}{R}}
   \right) \ud t^2\nonumber \\
& + \dfrac{\ud R^2}{1-\dfrac{2M(t)}{R}}-\dfrac{2H(t)R\, \ud t\ud R}
  {1-\dfrac{2M(t)}{R}}+ R^2 \ud \Omega^2,\label{eq:thakurta-areal}
\end{eqnarray}
where $H=H(t) = \dot a(t)/a(t)$ is the Hubble factor,
with $\dot a(t)$ standing for the derivative of $a(t)$ with respect to the 
cosmological time $t$, and we defined
\begin{equation}
    M= M(t)=m\,a(t). \label{M(t)}                
\end{equation}

It is interesting to compare this metric with other proposals in the 
literature. For instance, in Refs. \cite{Guariento:2012ri, daSilva:2015mja} 
the spacetime metric is given by a generalization of the McVittie solution 
\cite{McVittie:1933zz}, by letting the mass parameter be a function of the 
time coordinate. Indeed, the metric of Eq.~\eqref{eq:thakurta} can be 
characterized as such a generalized McVittie spacetime with a time-dependent 
mass parameter $m(t) = m a(t)$. This can be verified by  direct comparison 
between our Eq.~\eqref{eq:thakurta-areal} and Eq.~(9) shown in Ref. 
\cite{Guariento:2012ri}. However, the analyses made and the results presented 
in Refs. \cite{Guariento:2012ri, daSilva:2015mja} do not hold integrally for 
the Thakurta solution we are studying here, because a major part of them 
relied on the hypothesis that ${\dot{m}(t)}/{m(t)} < {\dot{a}(t)}/{a(t)}$, 
while for Thakurta both of such quantities are equal to each other.

Some aspects of the metric~\eqref{eq:thakurta} were analyzed in 
Ref.~\cite{Faraoni2009} (see also \cite{Faraoni:2008tx}). Note, however, that 
there has been some confusion 
concerning the nonrotating Thakurta metric and the Sultana-Dyer 
metric~\cite{sultana}, as pointed out in Ref. \cite{Carrera:2009ve}. Both are 
conformal to the Schwarzschild metric, but they are not the same. The 
difference between them lies in the dependence of the conformal factor $a^2$ 
as a function of the coordinates: in the former it is a function of the 
\emph{conformal time} $\eta$ alone, while in the latter it is a function of 
the \emph{Eddington-Finkelstein advanced time} $u= \eta +2m\ln \left| r- 2m 
\right|$. This difference is made clear by writing both metrics in diagonal 
forms as in Eq.~\eqref{eq:thakurta}, when a comparison can easily be 
performed [see, e.g., Ref.~\cite{Majhi:2014hpa} and compare Eq. (2.3) of that 
paper to  our Eq.~\eqref{eq:thakurta}].
 
In Ref.~\cite{Faraoni2009} the existence of a singularity at $r=2m$ was 
discussed, as well as a partial analysis of the causal structure of the 
related spacetime, considering a scale factor corresponding to a universe 
filled by dust, given by $a(t)\sim  t^{2/3}$, was performed. Also, in 
\cite{Culetu:2013hva} a metric of the same type of \eqref{eq:thakurta} is 
considered and some of the results presented in Sec.~\ref{sec:thakurta} of 
the present work were found.  In the present work the analysis of the causal 
structure for the case $a(t)\sim t^{2/3}$ is completed, and several other 
forms of the scale factor $a(t)$ are considered. In the following we 
investigate the main physical and geometrical properties of the corresponding 
spacetimes. Namely, we investigate the possible matter sources, the global 
structure, and the causal properties of each spacetime considering different 
forms of the scale factor $a(t)$.

\subsection{The energy-momentum tensor and the energy conditions}
From Eq.~\eqref{eq:thakurta}, it follows that the nonzero components
of the energy-momentum tensor $T\indices{_{\mu}^{\nu}}$
are given by
\begin{equation}
\begin{split}\label{tem}
&T\indices{_{t}^{t}}=  -\dfrac{3 H^2(t)}{8\pi f(R)}, \\
& T\indices{_t^r}=
-\dfrac{f^2(r)}{a^2(t)}\, 
T\indices{_r^t}=\dfrac{M(t)\,H(t)}{4\pi R^2 a(t)}\,, \\ 
& T\indices{_r^r}=T\indices{_\theta^\theta}=
T\indices{_\varphi^\varphi}= -\dfrac{3H^2(t)+2\dot H(t) } {8\pi f(R)}
\, ,
\end{split}
\end{equation}
where we defined 
\begin{equation}
f= f(R) = f(r) = 1 - \dfrac{2M}{R}=1 - \dfrac{2m}{r}. \label{eq:f(r)}
\end{equation}

With relations of Eq.~\eqref{tem} in hand, we can define the kinematic
quantities associated to the energy momentum tensor of the source, which
can be modeled as an imperfect isotropic fluid.
The flow vector $v_\mu$ is given by 
$$v_{\mu}=\left(-\sqrt{f(R)},\,0,\,0,\,0\right),$$ 
while the orthogonal projector $h_{\mu\nu}$ reads
$$h_{\mu\nu}= \text{diag}\left(0,a^2(t)\,f^{-1}(R),\,R^2,\,
R^2\sin^2\theta\right).$$
Using the standard definitions for the energy density $\rho$ and pressure
$p$ we find 
\begin{eqnarray}\label{rho}
\rho\equiv T_{\mu\nu}v^{\mu}v^{\nu}=\frac{3H^2(t)}{8\pi f(R)} \,,
\end{eqnarray}
 and
\begin{eqnarray}\label{p}
p\equiv \frac{1}{3}T_{\mu\nu}h^{\mu\nu}= -
\frac{3{H}^2(t)+2\dot{H}(t)}{8\pi f(R)} \,.
\end{eqnarray}
Also, since the energy-momentum tensor is not diagonal in the $tr$ sub-space, 
 there is a heat flow in the radial direction, whose energy flux is given by
\begin{eqnarray}\label{q}
q_{\sigma}=-T_{\mu\nu}v^{\mu}h\indices{^{\nu}_{\sigma}}= 
\left(0,\,-\frac{M(t)\,H(t)}{4\pi\, a(t)\,R^2f^{3/2}(R)},\,0,\,0\right).
\end{eqnarray}

Therefore, as the source for the rotating Thakurta spacetime 
\cite{Thakurta1981}, a possible source of the nonrotating Thakurta geometry 
given by Eq.~\eqref{eq:thakurta} is an isotropic fluid with a heat flow in 
the radial direction. The fluid quantities reduce to those of a homogeneous 
perfect fluid for large radial coordinate values, as expected. On the other 
hand, they may diverge at $R=2M$ which would imply a curvature singularity. 
This is probably the simplest source for this geometry. However, there are 
other possible sources, such as a mixture of a perfect fluid and a null 
fluid, but we do not consider these other more general sources here.

For completeness we state here the energy condition for a fluid with nonzero
energy flux. The relevant energy conditions for the present analysis are the
null (NEC), weak (WEC) and strong (SEC) energy conditions. Then we have 
(see, e.g., Ref.~\cite{Kolassis:1988})
\begin{align}
& {\rm NEC:}\;\;  \rho +p \geq 2|q|, \label{nec}\\
 & {\rm WEC:}\;\;  \rho -p + \Delta \geq 0,\quad  \rho +p \geq 2|q|,
 \label{wec}\\
& {\rm SEC:}\; \;  \rho +p \geq 2|q|, \quad 2p +\Delta \geq 0, \label{sec}
\end{align}
where $|q|= |q_\mu q^\mu|$, and 
\begin{equation}
 \Delta = \sqrt{\left(\rho+p\right)^2 - 4 |q|^2}.
\end{equation}
Using metric~\eqref{eq:thakurta} and the energy flux vector 
from Eq.~\eqref{q} it gives,
\begin{equation} \label{|q|}
 |q| = \frac{M(t)\,H(t)}{4\pi\, R^2\, f(R)}.
\end{equation}
In the analysis of the energy conditions given below, an interesting quantity
is the ratio $n(t,r)$ defined by
\begin{equation} \label{eq:ratio}
 n(t,r) = \frac{\rho +p}{2|q|} = \frac{-\dot H(t)\, R^2}{2M\, H(t)}.
\end{equation}
As seen from the above conditions, a sufficient condition to satisfy, for 
instance, the NEC is  $n(r,t) \geq 1$. Such a quantity shall be used to 
analyze some of the properties of the trapping horizons in the cases 
presented below.

\subsection{The Misner-Sharp mass} 

The Misner-Sharp mass is a measure of the gravitational active mass contained 
in a given volume of the spacetime (see, e.g., \cite{Hayward:1993ph}), and 
may furnish information  about the kind of objects enclosed in such a volume. 
In terms of the areal radius, $R=a(t)r$, the Misner-Sharp mass inside a 
closed surface of radius $R$ is defined by \cite{Hayward:1994bu}
\begin{equation}\label{MSmass}
M_{MS}=\dfrac{R}{2}\left( 1-\|\nabla 
R\|^2\right).
\end{equation}
For the Thakurta metric Eq.~\eqref{eq:thakurta-areal}, it follows
\begin{eqnarray}
\|\nabla R\|^2 
= -H^2(t)R^2\left(1-\frac{2M}{R}\right)^{-1}+1-\frac{2M}{R}.
\end{eqnarray}
Thus, substituting the last result into Eq.~\eqref{MSmass}
it gives
\begin{eqnarray}\label{MSmassf}
M_{MS}=M+\frac{H^2(t)R^3}{2\left(1-\dfrac{2M}{R}\right)}.
\end{eqnarray}
Notice that this expression contains a single concentrate mass 
contribution, the position-independent term $M$, which is compatible with the 
presence of a central object, and another contribution which depends on the 
radial coordinate, and that can be though of as the energy related to the 
fluid that fulfills the 
spacetime, since that second term grows with the Hubble factor. 
 This second contribution resembles the McVittie case \cite{McVittie:1933zz}, 
with the important difference that in McVittie spacetime the energy density 
is a homogeneous function.

In order to make this argument more rigorous we can use the decomposition of 
the Misner-Sharp mass into its Ricci and Weyl parts, $E_{R}$ and $E_{W}$, 
where the latter is interpreted as the source of the Coulombian part of the 
gravitational field. Following, e.g., Appendix D of 
Ref.~\cite{carrera-rmp-2010}, we obtain the following relation valid for 
spherically symmetric spacetimes:
\begin{gather}
 W^{\alpha \beta \mu \nu} W_{\alpha \beta \mu \nu} = \frac{48 E_{W}^2}{R^6} ,
 \label{eq:MS-Weyl}
\end{gather}
\noindent where $W$ is the Weyl tensor.
Applying Eq.~\eqref{eq:MS-Weyl} to the line element \eqref{eq:thakurta} we 
obtain 
\begin{gather}
 E_{W} = M= m\, a(t), \label{eq:mass}
\end{gather} 
\noindent
in agreement with our interpretation.
 
\subsection{The scale factor}

In the original Thakurta model \cite{Thakurta1981} the scale factor was not 
specified, even though it was assumed implicitly that it should describe 
an expanding cosmological model.
Here we keep such an assumption, by imposing that the $a(t)$ function
implies a big-bang type expanding model. 
However, since the mass of the central object increases with $a(t)$, 
cf. Eq.~\eqref{eq:mass}, it is interesting to modify its asymptotic behavior
at late times so that the final mass is finite, a necessary condition to
 have a black-hole type solution. In view of this,
we list here the main assumptions on $a(t)$.  
\begin{enumerate}[(1)]
 \item Big-bang hypothesis: $\lim_{t\to 0} a(t) =0$, with monotonically 
 increasing positive $a(t)$. 
 \item Expanding hypothesis: $\dot a(t)/a(t) \geq 0$.
 \item Cosmologically inspired (unbounded) models:
 \begin{enumerate}
  \item Asymptotically de Sitter cosmological model: $a(t)\sim e^{H_0 t}$ for
 large $t$.
  \item Asymptotically CDM model: $a(t) \sim t^{2/3}$.
  \item Alternative models as stiff matter and cosmic strings fluids, 
  $a(t)\sim t^\alpha$, $\alpha= 1/3$ and $\alpha = 1$.
 \end{enumerate}
 \item Bounded models $a(t)\sim$constant at large $t$.
 \begin{enumerate}
  \item Models with $\dot a(t)\gtrsim e^{-t/\tau}$, $\tau>2m$,  at late times
  --  no black-hole solutions.
  \item Models with $\dot a(t) \lesssim e^{-t/\tau}$,  $\tau<2m$, large times
  -- black-hole solutions.
 \end{enumerate}

\end{enumerate}

\section{Structural Analysis} \label{sec:structural}

\subsection{The curvature singularities}

First we analyze all the possible candidates for curvature singularities
that might
be present in metric~\eqref{eq:thakurta}.
Similarly to the Schwarzschild case, the metric is ill defined
at $r=2m$ ($R=2M$) and $r=0$ ($R=0$). 
Moreover, the points where $a(t)$ vanishes also may 
be singular points, and a deeper analysis is necessary in order to 
distinguish between curvature and coordinate singularities.

We start by analyzing the Ricci scalar which is given by 
\begin{eqnarray}
\label{eq:ricci}
\mathcal{R}=R\indices{_{\mu}^{\mu}}=\frac{6}{f(R)}\left(2H^2(t)+ 
\dot{H}(t)\right) .
\end{eqnarray}
Since $f(R)$ vanishes at $R=2M$, it follows that there is a curvature 
singularity at this point unless the factor between parenthesis, $2H^2(t)+ 
\dot{H}(t)$, vanishes as well. Then, the nature of the metric singularity  
$R=2M$ depends on the explicit form of the scale factor $H(t)$. We 
investigate this point in Sec.~\ref{sec:r=2m}. Additionally, the Ricci scalar 
is singular at points where $H(t)$, and/or $\dot H(t)$ diverge. This happens, 
for instance, when the factor $a(t)$ vanishes at a given time, as in the 
case of big-bang cosmological models. 
On the other hand, the Ricci scalar is nonsingular at $R=0$, and, hence, 
the behavior of the other curvature scalars at that point has to be 
analyzed.   

The Kretschmann scalar $\mathcal{K}$ is
\begin{equation}
\begin{split}
 \mathcal{K}=\,&R^{\alpha\beta\gamma\delta}R_{\alpha\beta\gamma\delta} 
= \frac{12}{f^2(R)}\left(H^4+\left(H^2+\dot H\right)^2\right)\\
& + \frac{16M^2}{R^4}\left(\frac{3}{R^2}-\frac{H^2}{f^2(R)}\right) 
,\label{eq:kret}
\end{split}
\end{equation}
showing that $R=0$ is a curvature singularity. Recalling that $R= ra(t)$,
this includes the locus where $r=0$ and the points where $a(t)=0$, 
i.e., at the initial time in a big-bang cosmological scenario
(where also $H(t),\;\dot H(t) \to \infty$).
Moreover, it results that $R=2M$ is a curvature singularity whenever
$H(t)\neq 0$ and $\dot H(t)\neq 0$.

\subsection{Trapping horizons}

Following Ref.~\cite{Hayward:1993mw}, we classify any sphere of symmetry $S$ of 
the spacetime according to the expansions $\Theta_{\pm}$ of the null 
congruences defined by outgoing and ingoing null rays, as follows, 
\begin{enumerate}[(1)]
\item if $\Theta_{+} \Theta_{-} < 0$,  we say that $S$ is \emph{regular};
\item if $\Theta_{+} \Theta_{-} > 0$, with $\Theta_{\pm} < 0$, we say that $S$ 
is \emph{trapped}; 
\item if $\Theta_{+} \Theta_{-} > 0$, with $\Theta_{\pm} > 0$, we say that $S$ 
is \emph{antitrapped};
\item if $\Theta_{+} \Theta_{-} =0$, then we say that $S$ is \emph{marginal}.
\end{enumerate}

This classification divides the spacetime in regions composed by each type of 
sphere. When drawing figures representing spacetime diagrams we indicate 
the different regions using the signs of 
expansions between parenthesis. Namely, we use ($- -$) to indicate trapped  
regions, ($+ -$) to indicate regular regions, and ($+ +$) to indicate 
antitrapped regions.

A \emph{trapping horizon} is defined as the tube foliated by  marginal spheres. 
In the case of a spherically symmetric metric such as 
Eq.~\eqref{eq:thakurta-areal}, these are given by the solutions of the 
equations
\begin{gather}
\Theta_{\pm} = \frac{2}{R} k^a_{\pm} \nabla_a R = 0 \,,
\end{gather}
and we get,
\begin{gather}
  H(t) R_\pm(t) \pm \left(1- \frac{2M(t)}{R_\pm(t)}\right) = 0,
\label{appR0}
\end{gather}
where the $\pm$ sign refers to outgoing (ingoing) null geodesics.

Taking $H(t) > 0$ it follows from Eq.~\eqref{appR0} that in the $R > 2M(t)$
region there are real solutions only for the ingoing expansion (minus 
sign) equation. 
Conversely, for $R<2M(t)$ the real solutions only exist for the outgoing 
expansion (plus sign) equation. 
Thus, only ingoing geodesics have vanishing expansion in the $R>2M(t)$ 
region, which is the region of interest for the present analysis.
The equation for the expansion of the ingoing null geodesics 
then yields two solutions for the trapping horizons, 
that we denote with an overhead hat throughout this paper,
\begin{gather}
  \hat{R}_\pm(t) =  \frac{1}{2 H(t)}\left(1 \pm \sqrt{1-
8M(t)\,H(t)}\right).
\label{appR}
\end{gather}

The subsequent study of geodesic completeness and of causal structure 
is simplified by using the original coordinates, i.e.,
in terms of the comoving radial coordinate $r$, and then it is useful to
write the relations defining the trapping horizons in terms of $r$.
Using the relations $R=ra(t)$ and $M(t)=m\, a(t)$, Eq.~\eqref{appR} can be 
cast into the form 
\begin{gather} \label{appr}
 \hat{r}_\pm = \frac{1}{2\dot{a}}\left(1 \pm \sqrt{1- 8m\dot{a}}\right)
 = \frac{4m}{1 \mp \sqrt{1 - 8m\dot{a}}}\,.
\end{gather}
It is seen that $\hat r_\pm$ are always larger than $2m$,
$\hat r_\pm> 2m$ for $\dot a\neq 0$. In the case $\dot a$ vanishes the 
horizon $\hat r_ -$ satisfies the equality $\hat r_ -=2m$ while $\hat r_+$
reaches arbitrarily large values. Here we refer to the loci defined by the
images  of the curves $\hat{r}_{\pm}(t)$ as
the outer ($+$) and the inner ($-$) horizons, respectively, since
$\hat{r}_{-}(t) \leq \hat{r}_{+}(t)\,, \, \forall\, t > 0$ for which the
trapping horizons exist.

Notice that both horizons only exist as long as $\dot{a} < {1}/{8m}$,
which provides very different structures depending on the behavior of the
scale factor $a(t)$.
Therefore, the trapping horizons can be roughly traced to the behavior of
$a$, that we divide into three broad classes\footnote{{Here, we choose
not to consider cases where $\dot{a}(t)$ oscillates between values below and
above $1/8m$. In those cases the trapping horizons could form an arbitrary
number of regular bubbles dividing the spacetime in an arbitrary number of
regions.}}: 

\begin{enumerate}[(a)]
 \item \label{caso-closed}For large $t$, $\dot{a}> 1/8m$, then the trapping
horizons have a finite life span, forming a closed curve and defining a bounded 
region in the  $(t,\, r)$ plane in its interior. For instance, considering the 
scale factor $a=\sinh^{2/3}(3H_0t/2)$, as an asymptotic $\Lambda$CDM model, as 
showed in Fig.~\ref{fig:horizonsLCDM}.

\begin{figure}[htb!]
\includegraphics[width=0.40\textwidth]{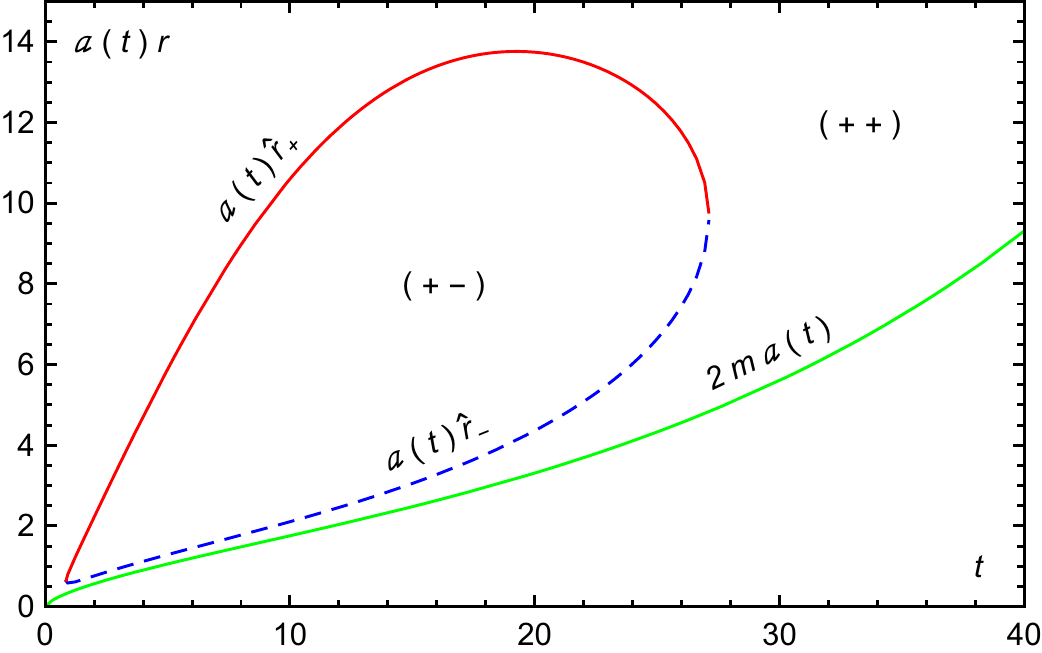}
 \caption{The outer [$\hat{R}_+=a(t)\hat{r}_+$] and inner 
[$\hat{R}_-=a(t)\hat{r}_-$] horizons represented
by red (solid) and blue (dashed) lines, respectively, as given by
Eq.~\eqref{appR}. Here the scale factor is $a(t)=\sinh^{2/3}(3H_0t/2)$,
with $H_0=0.05$, and the mass parameter was set to unity, $m=1$. The curve
$R=2m\, a(t)$ is also represented by the lowermost green (solid) line. }
\label{fig:horizonsLCDM}
\end{figure}

 \item \label{caso-open} For large $t$, $\dot{a} < 1/8m$, then the trapping
horizons are formed some time after the big-bang and remain for arbitrarily 
large $t$, dividing first quadrant of the 
$(t,r)$ plan in two open, unbounded regions.  For instance, considering the 
scalar factor $a=(t/t_0)^{2/3}$, as in the dust cosmological model, as showed 
in 
Fig.~\ref{fig:horizonsdust}.

\begin{figure}[htb!]
\includegraphics[width=0.40\textwidth]{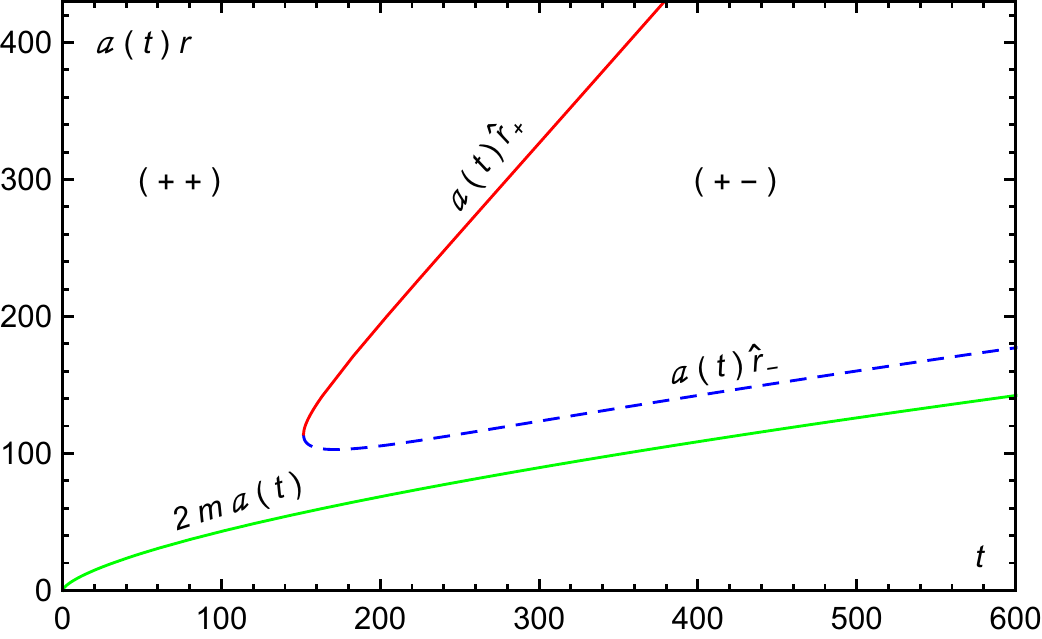}
 \caption{The outer [$\hat{R}_+=a(t)\hat{r}_+$] and inner 
[$\hat{R}_-=a(t)\hat{r}_-$] horizons represented
by red (solid) and blue (dashed) lines, respectively, as given by
Eq.~\eqref{appR}. Here the scale factor is  $a(t) = (t/t_0)^{2/3}$, 
with $t_0=1$, and the mass
parameter was set to unity, $m=1$. The curve $R=2m\, a(t)$ is also
represented by the lowermost green (solid) line. }
 \label{fig:horizonsdust}
 \end{figure}
 
 \item \label{caso-open-zero} For large $t$, $\dot{a} \leq 1/8m$ and, 
moreover, $\dot{a} \leq 1/8m$ at initial times, then the trapping horizons 
are formed at the big-bang and remain for arbitrarily large $t$, dividing 
the 
first quadrant of the $(t,\,r)$ plan into three open, unbounded regions.  For 
instance, considering the scalar factor $a= \tanh(t/t_0)$, with $t_0\geq  8m$ 
as showed in Fig.~\ref{fig:horizonsTanh}. In the case $\dot{a} = 1/8m$ for 
all times the two trapping horizons coincide. This is a particular 
case of linear scale factors $a(t)\sim t$ which are analyzed in the next 
section.

\begin{figure}[htb!]
\includegraphics[width=0.40\textwidth]{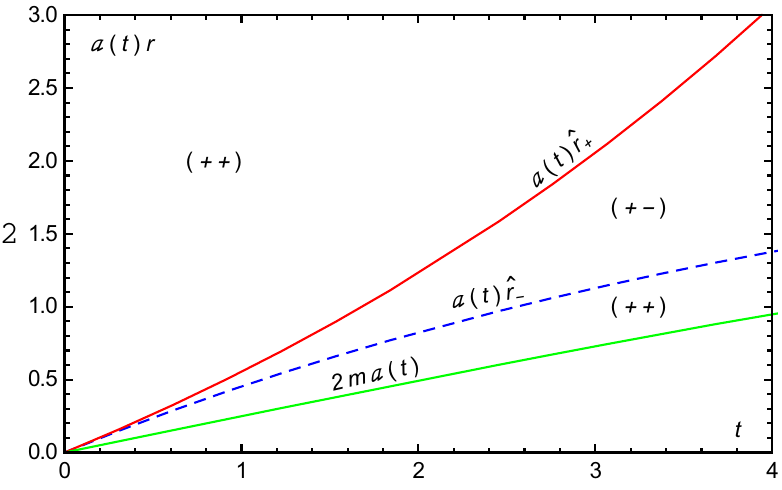} 
\caption{The outer 
[$\hat{R}_+=a(t)\hat{r}_+$] and inner [$\hat{R}_-=a(t)\hat{r}_-$] horizons 
represented by red (solid) and blue (dashed) lines, respectively, as given 
by Eq.~\eqref{appR}. Here the scale factor is  $a(t) = \tanh(t/t_0)$, with 
$t_0 = 8m$, and the mass parameter was set to $m=1.25$. The curve $R=2m\, 
a(t)$ is also represented by the lowermost green (solid) line. }
 \label{fig:horizonsTanh}
 \end{figure}
 
 \end{enumerate}

\subsection{Geodesic completeness} \label{sec:geo-complete}

In order to characterize the spacetime described by the Thakurta
metric, we need to investigate the characteristics of the patch covered by
the given coordinates. A very important property that we must verify is if
the limits of the region covered by the chosen coordinates are at finite or
infinite proper distance of the events at finite coordinates. In other words,
we must check the geodesic completeness. For that 
we study the asymptotic behavior of the affine parameter as the time coordinate
$t$ tends to infinity (future behavior), and to zero (past behavior)
along the null geodesics.

\subsubsection{Future behavior of null geodesics}

The outgoing null geodesics are trivially future complete ($t \to \infty$), 
since the spacetime is asymptotically FLRW as $r \to \infty$ and its results 
can be used. Therefore, we have to verify completeness for ingoing null 
geodesics, which approach $r =2m$. For that we consider the equation for 
ingoing null geodesics in terms of the affine parameter  $\lambda$,
\begin{gather}
 r^\prime_- = - \frac{t^\prime}{a} \left(1 - \frac{2m}{r_-} \right)
\,,\label{drdt}\\
 t'' = \left(\frac{2m}{ar_-^2} - H\right)t'^2 \, ,\label{dtdlambda}
\end{gather}
\noindent
where the prime stands for the 
$\lambda$-derivative. 

Equation~\eqref{drdt} can be rewritten in terms of the conformal time 
$\ud\eta =
\ud t/a$ and integrated to produce 
\begin{gather}
 r_- + 2m \ln (r_- - 2m) = -\eta +C\,, \label{r-solution}
\end{gather}
\noindent
where $C$ is an arbitrary constant.

In order to analyze geodesic completeness, we study the behavior of the affine 
parameter as $t$ tends to infinity along ingoing null geodesics. Using 
Eq.~\eqref{dtdlambda} we obtain for large $t$,
\begin{gather}
\label{eq:lambda1}
\frac{t''}{t'} \approx K \frac{t'}{a} - \frac{\dot{a} t'}{a},  
\end{gather}
where we defined $K\equiv {2m}/{r_\infty^2}$, with $r_\infty$ denoting 
the value of $r_-(t)$ in the limit of large times.
\noindent
Integrating Eq.~(\ref{eq:lambda1}) we have
\begin{gather}
\label{eq:lambda2}
 \ln t' \approx  K \eta(t) - \ln a + \text{constant}. 
\end{gather}
Then the parameter $\lambda$ is given by the integration of 
Eq.~(\ref{eq:lambda2}) as 
\begin{gather}
 \lambda(t) \sim \int^t e^{-K\eta(u)} a(u) \ud u\,,\label{eq:lambda3}
\end{gather}
\noindent 
and the convergence or divergence of the last integral as $t\to\infty$ 
determines if the spacetime is geodesically incomplete or complete, 
respectively. A more cautious proof of this result, considering the effect of 
subleading terms, is the subject of Appendix~\ref{app:subleading}.

The convergence of the integral in Eq.~\eqref{eq:lambda3} depends on $\eta$,
and we must consider the two aforementioned possibilities.

\begin{enumerate}[(i)]
\item \label{eta-bounded} $\eta(t)$ is bounded as $t \to \infty$.
 
\item \label{eta-unbounded} $\eta(t)$  is unbounded as $t \to \infty$. 
\end{enumerate}

\begin{figure}[htb!]
\includegraphics[width=0.4\textwidth]{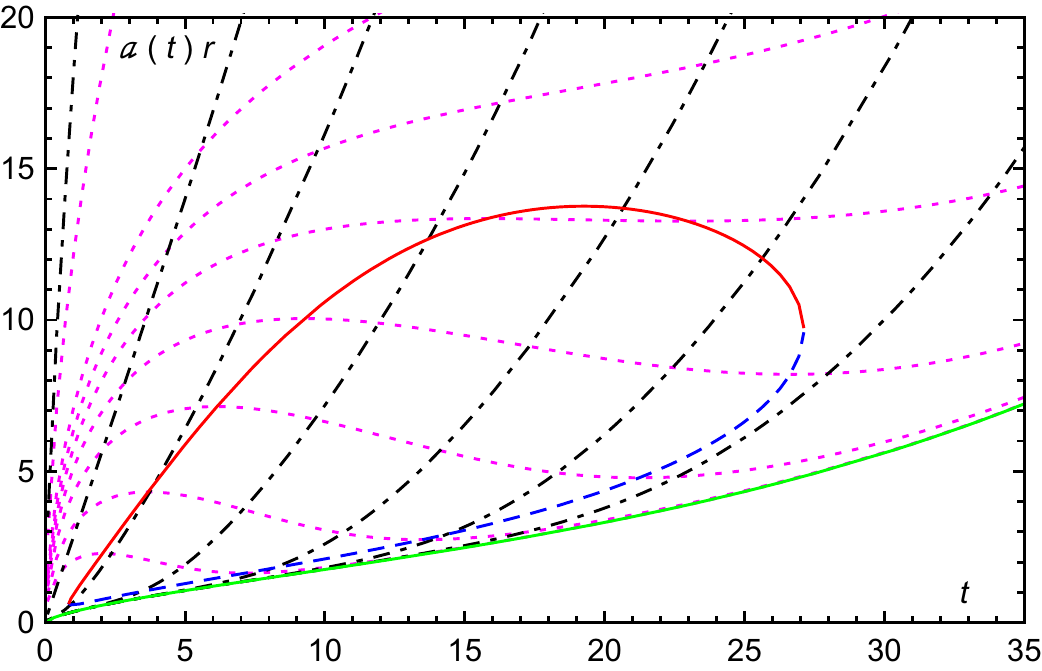}
 \caption{Ingoing and outgoing geodesics represented by magenta (dotted) and
black (dotted-dashed) lines, respectively. The scale factor
is $a(t)=\sinh^{2/3}(3H_0 t/2)$, with $H_0=0.05$. The
outer/inner horizons and the $r=2m$ curve are represented respectively by
the blue (solid), red (dashed) and  green (solid, lowermost) lines. The mass 
parameter $m$ has been set to unity.}
  \label{fig:geodesicsLCDM}
\end{figure}

If $\eta$ is bounded, case~\ref{eta-bounded}, Eq.~\eqref{r-solution} implies
that the surface $r=2m$ is unreachable by ingoing null geodesics. This
behavior can be seen in Fig.~\ref{fig:geodesicsLCDM} as, for example,
taking $a(t)=\sinh^{2/3}(3H_0t/2)$.  Thus, defining $\eta_{\infty} = \lim_{t
\to \infty} \eta(t)$ we can define $r_\infty \equiv r_-(\eta_\infty) > 2m$.
In terms of the areal radius we have $R_-(t) \sim a(t) r_\infty$ for large
$t$, which implies that the expansion of ingoing geodesics is positive in
this limit. 

Thus, we have that $\eta(t) < \eta_\infty$, for all $t>0$, such that
\begin{gather}
 \lambda \sim \int^\infty e^{-K\eta(u)} a(u) \ud u > e^{-K\eta_\infty} 
 \int^\infty a(u) \ud u \to \infty\,,
\end{gather}
hence, if $\eta(t)$ is bounded, the Thakurta spacetime is future geodesically 
complete.  

On the other hand, in case~\ref{eta-unbounded}, the limit of the null ingoing 
geodesics as $t \to \infty$ is $r_\infty = 2m$. In order to address this case 
we analyze the particular situation where the scale factor is asymptotically 
linear, $a(t) \sim a_0 t$ for large $t$. 
In this case we have, for large $t$,
\begin{gather}
 \lambda \sim \int^\infty \exp[{-{K \ln(u)}/{a_0}}]\, u \ud u = 
 \int \frac{1}{u^{{K}/{a_0}-1}} \ud u\,.
\end{gather}
Therefore, the convergence depends on the difference ${K}/{a_0}-1 = 
{1}/{(2a_0m)}-1$. Moreover, since $K\equiv2m/r^2_{\infty}$ we obtain
\begin{gather}
 \frac{1}{2a_0m}-1 \leq 1 \Leftrightarrow  a_0 m \geq 1/4 
 \Rightarrow \lambda \to \infty \quad \text{(complete)}\,,\\
  \frac{1}{2a_0m}-1 > 1 \Leftrightarrow  a_0 m < 1/4 
  \Rightarrow \lambda < \infty\,\quad \text{(incomplete)}.
\end{gather}

\begin{figure}[htb!]
\includegraphics[width=0.4\textwidth]{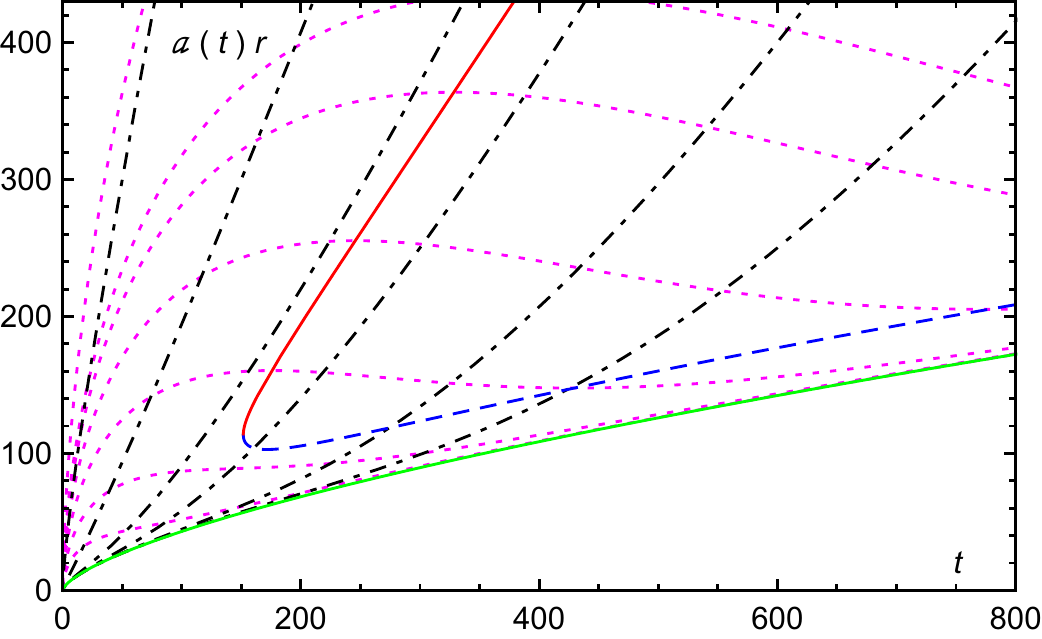}
 \caption{Ingoing and outgoing geodesics represented by magenta (dotted) and
  black(dotted-dashed) lines, respectively, for the case 
$a(t)=(t/t_0)^{2/3}$, with $t_0=1$.
The outer/inner horizons and the $R=2m\, a(t)$ curve are represented
respectively by the blue (solid), red (dashed) and  green (solid, lowermost)
lines. The mass parameter $m$ has been put to unity.}
 \label{fig:geodesicsdust}
 \end{figure}

By extension, we conclude that in case~\ref{eta-unbounded}, the Thakurta
spacetime is geodesically complete if, for large $t$, $a(t) > {t}/{(4m)}$
and geodesically incomplete if $a(t) < {t}/{(4m)}$. In particular, this
implies that all Thakurta spacetimes with sublinear $a(t)$ are geodesically
incomplete, as for example $a(t)=(t/t_0)^{2/3}$ showed in
Fig.~\ref{fig:geodesicsdust}.

 \subsubsection{Past behavior of null geodesics}
 
 Here we show that the singularity $a(t) = 0$ is always at a finite time in 
the past for all models we consider in this work. For this we study the behavior 
of the null geodesics in the limit $t\to 0$.

 For ingoing null geodesics we must consider Eqs.~\eqref{dtdlambda} 
and~\eqref{r-solution}. Following a similar reasoning used in the study of
the future behavior of null geodesics,
we find that if $\eta$ is bounded from below then $r_-(t)$ is 
bounded, i.e.,  $2m < r_-(t) < r_{max}$, while if $\eta$ is unbounded $r_-(t) 
\to \infty$ as $a(t) \to 0$, which means that, by Eq.~\eqref{r-solution}, we 
can use the approximation $r_-(\eta) \sim -\eta$. 

First, we consider the bounded case, for which $r \to r_{max}$ as $t 
\to 0$. We can choose $\eta( t = 0) = 0$ with no loss of generality. 
The procedure is analogous to the one used to investigate completeness to the
future, since the behavior is governed by Eq.~\eqref{dtdlambda}. Redefining
$K = {2m}\big/{r^2_{max}}> 0$ in Eq.~\eqref{eq:lambda1} and integrating, 
we obtain  
\begin{gather}
\Delta \lambda \approx \int_t^0 \exp  \left[- K
\eta(u)\right]a(u) \ud u\,. \label{eq:lambda-past-ingoing}
\end{gather}
Since $\eta(t) \geq 0$ and $K> 0$,  it follows 
$ \exp  \left[- K
\eta(u)\right] \leq 1$, and hence
\begin{gather}
| \Delta \lambda | \leq \left| \int_t^0 a(u)\ud u \right|,
\end{gather}
\noindent
which is always finite. 

Similarly, if $\eta$ is unbounded from below, $\lim_{t \to 0} 
\eta(t) = - \infty$, we obtain, instead of 
Eq.~\eqref{eq:lambda-past-ingoing}, the following relation,
\begin{gather}
  \Delta \lambda \approx \int_t^0 \exp  \left[\frac{2m 
}{\eta(u)}\right] a(u) \ud u\,,
\end{gather}
\noindent
which is always finite, since $\exp  \left[{2m }/{\eta(u)}\right]
\leq 1$ for $\eta(u) < 0$.

Therefore, all the ingoing null geodesics are 
incomplete to the past, which proves that the surface $a(t) =0$ is a 
singularity in the past of all events in Thakurta spacetimes, and it is 
justified to call it a big-bang singularity in this context.

Using similar arguments we can also prove that the outgoing null geodesics are 
incomplete to the past.

\subsection{Conformal boundaries}

In order to build the conformal diagrams of the Thakurta spacetimes, we have 
to study the properties of the conformal boundaries at the past $a(t) \to 0$ 
and  at the future $t \to \infty$. Using the conformal time coordinate $\eta = 
\displaystyle{\int \frac{\ud t}{a(t)}}$ and the tortoise coordinate $r^* = 
\displaystyle{\int \dfrac{\ud r}{f(r)}}$, the Thakurta metric can be written 
in the form
 \begin{gather}
  \ud s^2 = a^2(\eta) f(r)\left( - \ud \eta^2 + \ud r^{*2}
  + r^2 (r^*)\ud \Omega \right)\,, \label{eq:conformal}
 \end{gather}
which is conformally flat in the $(\eta, r^*)$ submanifold. Since $-\infty < 
r^* < \infty$, the behavior of $\eta$ as $t \to 0$ determines the properties of 
the $a(t) = 0$ surface in the past, as in the FLRW models \cite{wald, 
hawking}. 

\begin{enumerate}[(1)]
 \item If $\lim_{a(t) \to 0} \eta(t)$ is finite, then the Thakurta spacetime is 
conformally related to a portion of Minkowski spacetime at the future of a 
constant time spacelike hypersurface. This implies that the past conformal 
boundary is spacelike. 
 
 \item If $\lim_{a(t) \to 0} \eta(t) = -\infty$, then the past conformal 
boundary of the Thakurta spacetime corresponds to the boundary of the 
Minkowski spacetime and it is null. 
\end{enumerate}
 
Analogously, we have the same two cases in the limit $t \to \infty$ for the 
future boundary:
\begin{enumerate}[(1)]
 \item If $\lim_{t \to \infty} \eta(t)$ is finite, then the Thakurta spacetime 
is conformally related to a portion of the Minkowski spacetime at the past of a 
constant time spacelike surface. The future boundary is spacelike in 
this case, as in the particular sample depicted in 
Fig.~\ref{fig:diagrama-Sinh23}.
 
 \item If $\lim_{t \to \infty} \eta(t)=\infty$, then the future conformal 
boundary of the Thakurta spacetime corresponds to the boundary of the 
Minkowski spacetime and it is null. 
\end{enumerate}

Therefore, by choosing the asymptotic behavior of the function $a(t)$ in the 
future and in the past we can build Thakurta models with  the four possible 
combinations in terms of conformal boundaries.  We shall show below some 
examples of topologies given by different examples of $a(t)$.

\section{The surface at 
\texorpdfstring{$\boldsymbol{r = 2m}$}{Lg}}
\label{sec:r=2m}

In Sec.~\ref{sec:geo-complete}, we stated the conditions on $a(t)$ that imply 
the surface $r=2m$ is reached in a finite affine parameter by null
ingoing geodesics.
  
We need to evaluate the properties of that 
surface, namely, if it is singular or regular. Since the Ricci scalar is given 
by Eq.~\eqref{eq:ricci}, it may diverge as $t \to \infty$ and $ r \to 2m$. In 
order to determine whether that divergence of the Ricci scalar is physical or 
not, we evaluate its limit along some classes of physical trajectories.

\subsection{Null trajectories} \label{null-trajectories}

The null ingoing geodesics satisfy the relation given in 
Eq.~\eqref{r-solution}. 
Near the $r \to 2m$ limit it can be approximated by
\begin{gather}
 r_- \approx 2m + C'\exp\left[\dfrac{-\eta}{2m}\right]\,,
\end{gather}
\noindent
which leads to
\begin{gather}
f[r_-(\eta)] \sim \frac{C'}{2m}\exp\left[\dfrac{-\eta}{2m}\right]\,.
\end{gather}

Therefore, the Ricci scalar is given by
\begin{gather}
 \mathcal{R} \sim  \exp\left[\frac{\eta}{2m} \right] \times \frac{\dot{a}^2 
 + a\ddot{a}}{a^2} \,.
\end{gather}

Assuming that $a(t)$ behaves asymptotically as a power of $t$, $a(t)\sim 
t^\alpha$ with $\alpha>0$, it follows $a(\eta) \sim 
\eta^{{\alpha}/{(1-\alpha)}}$ and thus, if $\dot{a}^2 + a\ddot{a} \neq 0$, 
$\mathcal{R} \to \infty$ since the exponential dominates as $\eta \to \infty$. 
The form of the scale function that provides $\dot{a}^2 + a\ddot{a} = 0$ is 
$a(t) = c_1 t^{1/2}$. In this case, the singularity appears in the 
Kretschmann scalar \eqref{eq:kret}, which in the limit $r\to 2m$ goes as 
$
\mathcal{K}\sim  \exp\left[{\eta}/{m} \right] \times \eta^{-8}$
and that also diverges in the limit $\eta \to \infty$. 

The conclusion is that the surface $r=2m$ is singular for lightlike geodesics 
if the scale factor grows as a power law (or at a higher rate) at large 
cosmological times.

\subsection{Timelike observers}

Lightlike (null) trajectories are the fastest allowed paths for sign travel 
in general relativity, and hence it would be interesting to check if the 
analysis performed in
Sec.~\ref{null-trajectories} would give a different result if we used
timelike trajectories, that move at lower velocities.

Here it is convenient to use the conformal time $\eta$ as the time coordinate, 
bringing the line element to the form given in Eq.~\eqref{eq:thakurta}.

We can parametrize the 4-velocity of any timelike observer as
\begin{gather}
 u^0 = \frac{\ud \eta}{\ud s} = \frac{ \cosh \omega }{a\sqrt{f(r)}} ,
 \quad u^1 = \frac{\ud r}{\ud s} =\frac{\sqrt{f(r)} \sinh \omega}{a} ,
\end{gather}
\noindent
where $s$ is the proper time of the observer and $\omega$ is its 
\emph{rapidity}.
Therefore, we have
\begin{gather}
 \frac{\ud r}{\ud \eta} = \frac{u^1}{u^0} = f(r) \tanh \omega ,
\end{gather}
which gives 
\begin{gather}
 \eta(r) = \int \frac{\ud r}{f(r) \tanh \omega}.
\end{gather}

Since we are interested in ingoing observers that reach the $r=2m$ surface at 
a finite proper time, it is appropriate to take $\omega = - |\omega|$. In 
order to simplify the problem further, we consider that $\omega$ is constant 
along the trajectory. Then we obtain
\begin{gather}
 \eta = - \frac{1}{\tanh |\omega|} \left( r + 2m \ln\left(r-2m\right) \right)
  + \text{constant},
\end{gather}
and hence, close to the limit $r \to 2m$, it yields 
\begin{gather}
 r(\eta) \sim C \exp \left[- \frac{\eta \,\tanh |\omega|}{2m}\right] + 2m \,.
\end{gather}
Next we find the approximate form for the function $f(r)$ along a timelike 
geodesic in the limit $r \to 2m$,
\begin{gather}
 f[r(\eta)] \sim  \exp \left[- \frac{\eta \, \tanh |\omega| }{2m}\right].
\end{gather}
\noindent
From these results, and with the hypothesis that $a(t)$ behaves  
asymptotically as a power of $t$, it is seen that $f$ dominates the limit of 
$\mathcal{R}$ as $\eta \to \infty$,  irrespective of the value of $|\omega|$, 
and the timelike geodesics hit a singularity ($\mathcal{R} \to \infty$) at 
$r=2m$.

This proof also applies for ingoing observers with variable $\omega$ provided 
$|\omega|$ is bounded from below.

\subsection{Building Thakurta models with a nonsingular 
\texorpdfstring{$\boldsymbol{r=2m}\,\ $}~surface}
\label{sec:buildbh}

As seen from the previous analysis, the Thakurta metric does not
correspond to a black-hole spacetime if the scale factor follows the
standard cosmological scenario. Notice that in order to obtain a black-hole
model from that metric, it has to be associated to an incomplete patch of
the spacetime and the locus $r=2m$ must be a nonsingular surface,
which would be the expected locus of the event horizon.

The above analysis suggests that, as $f[r(t)]$ goes to zero exponentially
for timelike and null lines, the derivatives of $a(t)$ should also vanish
exponentially fast in a time scale shorter than that of $f[r(t)]$. This also
implies that $a(t)$ has to be bound as $t \to \infty$, and hence
all models where $a(t)$ is unbounded are singular at $r=2m$. Then we assume
here that $\lim_{t \to \infty} a(t) = a_{\infty} > 0$. In this case, the
conformal time behaves linearly with respect to the cosmic time, that is,
$\eta \sim t/a_{\infty}$ for large times.

Consider now the Kretschmann scalar, which is given by Eq.~\eqref{eq:kret}.
The dominant term in $\mathcal{K}$ in the limit $r\to 2m$ is 
${\dot{a}^2}/{f^2[r(t)]}$. 
Thus, in order to have a finite $\mathcal{K}$, assuming $\dot{a} \sim e^{- 
{t}/{\tau}}$, the constraint
\begin{gather}
 \frac{1}{2m} - \frac{2}{\tau} < 0 \Longleftrightarrow \tau < 4m,
\end{gather}
must be satisfied.
Therefore, a sufficient condition for the surface $r=2m$ being nonsingular 
is 
$\dot{a} < \mathcal{O}(e^{-t/4m})$. This condition may be fulfilled by models with 
$a(t) = \tanh({t/t_0})$ studied in Sec.~\ref{sec:causal}, if $t_0 = 2\tau < 
8m$. In cases like this the spacetime can be extended across the surface 
$r=2m$ and its analytical extension may be a black-hole or white-hole 
spacetime.

\section{Analysis of sources and causal structure} 
\label{sec:causal}

\subsection{Preliminaries}

Different models can be crafted for different choices of $a(t)$.
The main differences  between them are related to the existence or 
not of a singularity at $r=2m$, and the behavior of the
geodesics going to both time boundaries (past singularity and future time
infinity). 
We shall then present some examples of interesting cases with different causal 
structures. In all the examples discussed in this section the value of the 
mass parameter is fixed to $m=1$.

\subsection{Solutions with unbounded scale factor}
\label{sec:unbounded}

\subsubsection{Preliminaries}

Let us start with a few examples of scale factors that may describe some
phase of expanding cosmological models. In all of the cases 
studied in this section 
the function $a(t)$ starts with zero value at $t=0$, a big-bang type
singularity, and growths monotonically with the cosmological time.

\subsubsection{\texorpdfstring{$a(t)= 
\sinh^{3/2}(t/t_0)$}  {Lg}: A smooth \texorpdfstring{$\Lambda$}{Lg}CDM model}
\label{sec:LCDM}

The first case is a 
$\Lambda$CDM universe with $a(t)=\sinh^{2/3}(t/t_0)$, $t_0=2/(3H_0)$, as the 
model used in Ref.~\cite{Lake:2011ni}. In this case the 
central mass $M(t)=ma(t)$ grows exponentially fast at late times. The horizons 
exist only by a  finite amount of time and form a bubble that does not reach 
the future  infinity, as represented in Fig.~\ref{fig:horizonsLCDM}.

\begin{figure}[htb!]
 \includegraphics[width=.4\textwidth]{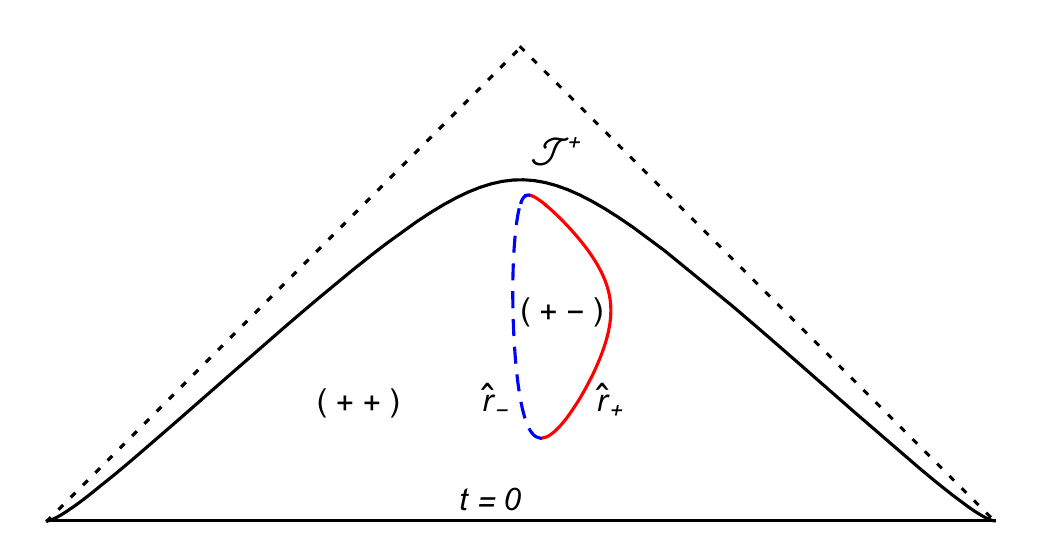}
 \caption{The causal diagram for a $\Lambda$CDM model where $a(t) = 
\sinh^{2/3}(3H_0t/2)$, with $mH_0 = 0.03$. There 
is a spacelike initial singularity at $t=0$. Also the conformal time $\eta$ is 
bounded 
and the future infinity is spacelike. The dotted lines are not part of  the 
spacetime and are drawn for comparison to the other situations.}
  \label{fig:diagrama-Sinh23}
\end{figure}

The causal diagram depicted in Fig.~\ref{fig:diagrama-Sinh23} summarizes the 
geometrical properties of the corresponding spacetime. 
The locus $r=2m$ is beyond the future infinity and the spacetime  
presents a bounded conformal time $\eta$, which is represented by the 
spacelike (black, solid) curve delimiting the spacetime.
The regular region lies inside the 
closed surface generated by the trapping horizons, the surface $t=0$ is 
spacelike (and singular), and the spacetime is future geodesically complete.
No black hole is present, and the shown diagram is in fact the maximal 
analytical extension. 

The choice $a(t)=\sinh^{2/3}(3H_0t/2)$ is motivated by the standard 
cosmological scenario, which has a phase dominated by cold dark matter 
(dust fluid, $p=0$) and approaches a de Sitter fluid ($p=-\rho$) at late times,
and has been used in Refs.~\cite{Lake:2011ni,daSilva:2012nh}. The related 
energy density and pressure [see Eqs.~\eqref{rho} and \eqref{p}] are given 
respectively by
 \begin{equation} \label{prhopSinh23}
   8\pi\, \rho = 
\dfrac{3H_0^2\coth^2\left[3H_0\, t/3\right]}{f(r)}, \quad 
8\pi\, p = -
\dfrac{3H_0^2}{f(r)},   
 \end{equation}
  where $f(r)$ is given by Eq.~\eqref{eq:f(r)}.
\begin{figure}[htb!]
 \includegraphics[width=.4\textwidth]{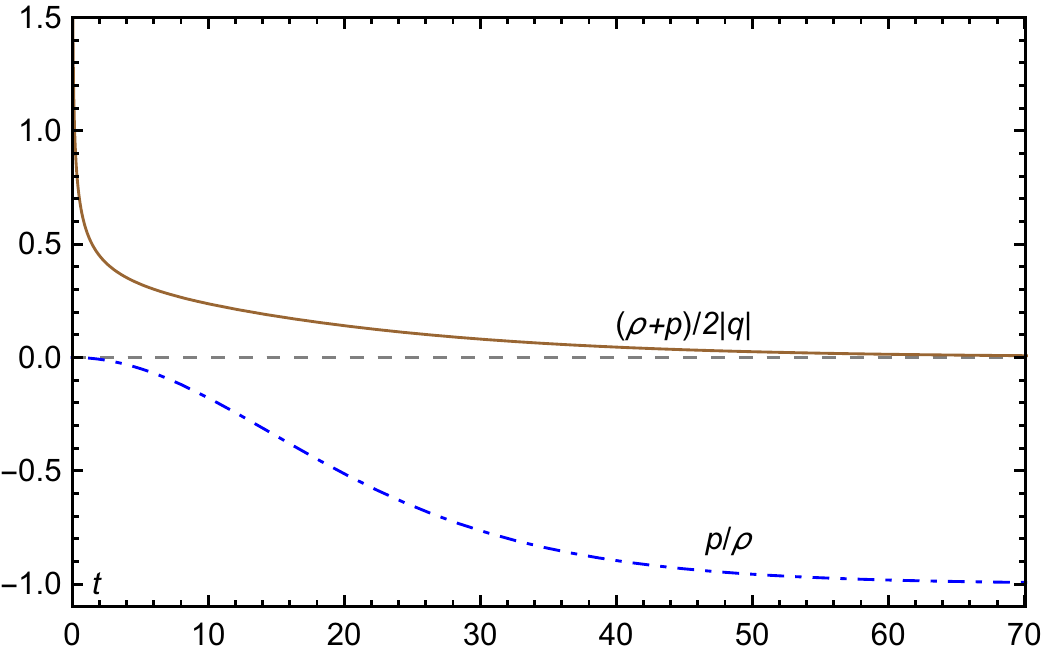}
 \caption{The curves for $w(t)=p/\rho$ and $(p+\rho)/2|q|$ 
 in the case $a(t)=\sinh^{2/3}(3H_0t/2)$, 
for $H_0=0.03$ and $r=3\times m$ (we have chosen $m=1$).}
  \label{fig:pOrho-Sinh23} 
\end{figure}
 
Since the pressure is time independent, the ratio $w(t)=p/\rho$, plotted in 
Fig.~\ref{fig:pOrho-Sinh23}, indicates the time evolution of the 
energy density. 
At early times the energy density is large, the fluid behaves like a dust 
fluid, while at late times the energy density is also a constant. The final 
state is $p=-\rho= -3H_0^2\Big/\!\!\left(1-\dfrac{2m}{r}\right)$.

The only nonzero component of the heat flow $q_\mu$ is the one along 
the radial direction, $q_r$.
Eq.~\eqref{|q|} then gives
\begin{equation} \label{qr1}
 |q| = \frac{2m H_0 \, \coth\left(3 H_0 t/2\right)}
{r^2\sinh^{5/3}\left(3 H_0 t/2\right) f(r)}, 
\end{equation}
 where $f(r)$ is given by Eq.~\eqref{eq:f(r)}. This energy flux decreases
 with $r^{-2}$, it is arbitrarily large at initial times and
 tends to zero at late times.  

The energy density and pressure, given by Eqs.~\eqref{prhopSinh23},
together with the energy flux~\eqref{qr1}, define the energy-momentum
content of the spacetime. Regarding to the energy conditions 
[cf. Eqs.~\eqref{nec}, \eqref{wec}, and \eqref{sec}] the important
quantity to analyze is ratio $n(t,r)=(\rho+p)/2|q|$, defined in
Eq.~\eqref{eq:ratio}, which in the present case is 
$n(t,r)= \left({3H_0r^2}\big/{2m}\right){\rm csch}\left(3H_0t\right)
\sinh^{2/3}\left(3H_0t/2\right)$, and so the energy conditions
strongly depend on time and radial coordinates. 
For small times it approaches to
$n(t,r) \sim 9r^2H_0^{2/3}/\left(4m\,t^{1/3}\right)$, while for large times it
results $n(t,r) \sim 3r^2H_0{\rm e}^{-H_0t}/\left(2m\right)$. 
Hence, at very initial times and finite $r> 2m$,
the constraint $\rho+p \geq 2|q|$ holds and it can be verified that all the
energy conditions are satisfied. Moreover, since (for fixed $r$) the
function $\rho +p$ vanishes faster than the energy flux $|q|$, at late times
(for fixed $r$) the constraint $\rho+p\geq 2|q|$
does not hold and the NEC, the WEC, and the SEC are violated.
Such a behavior is depicted in Fig.~\ref{fig:pOrho-Sinh23}.
On the other hand, since $|q|$ falls with $1/r^2$, for any fixed time, 
at sufficiently large radial coordinates ($r\rightarrow\infty$), 
the NEC, the WEC, and the SEC are satisfied.

We can also evaluate the energy conditions along the trapping horizons 
$\hat r_+(t)$ and $\hat r_-(t)$, and investigate the character of the
trapping horizons in the spirit of the classification and theorems of 
Refs.~\cite{Hayward:1993mw} and~\cite{Andersson:2005gq}. 
To do that properly, and to be able to compare the different cases, we
report on such a subject in Appendix \ref{app:horizonsclass}.

\subsubsection{\texorpdfstring{$a(t)=\left({t}/{t_0}\right)^{2/3}$} {Lg}: 
CDM (dust fluid) model} \label{sec:dust}

\begin{figure}[htb!]
 \includegraphics[width=.4\textwidth]{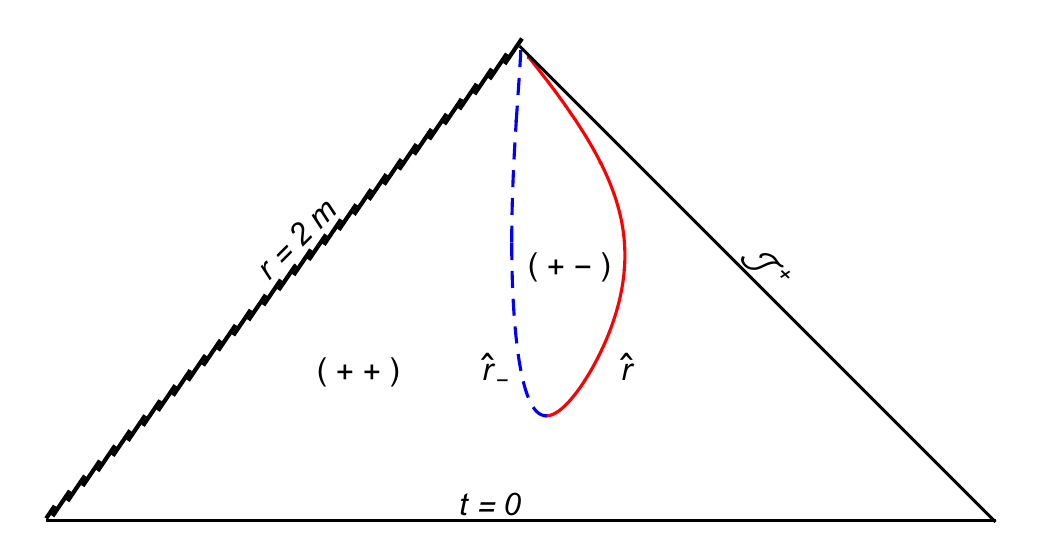}
 \caption{The causal diagram for the  dust model $a(t) = 
\left(t/t_0\right)^{2/3}$, with $t_0=1=m$. There is an initial 
singularity 
at $t=0$, a null singularity at $r=2m$. The trapping horizons are timelike for 
large times. }
 \label{fig:diagrama-dust}
\end{figure}

Another important example is a dust model with $a(t)=(t/t_0)^{2/3}$ for all 
$t>0$.
For this scale factor the trapping horizons have similar forms 
to those presented in Fig.~\ref{fig:horizonsdust}. Here the 
conformal time $\eta$ is unbounded and hence the lightlike geodesics reach 
the future null infinity. At late times both of the horizons are timelike and 
form a closed curve that reaches the timelike infinity. The causal diagram is 
given in Fig.~\ref{fig:diagrama-dust}.

Similarly to the case of the last section, the regular region is inside the 
closed curved formed by the trapping horizons. The surface $t=0$ is singular 
and spacelike. An important difference is that here the lightlike surface 
$r=2m$ belongs to the spacetime, it is a singular boundary of the spacetime. 
Moreover, the spacetime is geodesically incomplete also to the future, for 
geodesics hitting the singularity at $r=2m$.  The 
Fig.~\ref{fig:diagrama-dust} then shows the maximal analytical 
extension of the solution.

The $a(t) = \left(t/t_0\right)^{2/3}$ function produces a dust background 
fluid with a heat flow in the radial direction,
 \begin{equation} \label{prhodust}
 \begin{split}
   & 8\pi\, \rho = \dfrac{4}{3t^2} \dfrac{1}{ f(r)},\qquad
8\pi\, p = 0, \; \\
&  8\pi\,|q| =  \dfrac{4m\,t_0^{2/3}}{3t^{5/3}} \dfrac{1}{ r^2f(r)},
 \end{split}
 \end{equation}
 where $f(r)$ is given by Eq.~\eqref{eq:f(r)}.
Since the properties of this fluid are very simple we do not show the graphs 
of the fluid quantities nor of the ratio $p/\rho$ for this case.

\begin{figure}[htb!]
 \includegraphics[width=.4\textwidth]{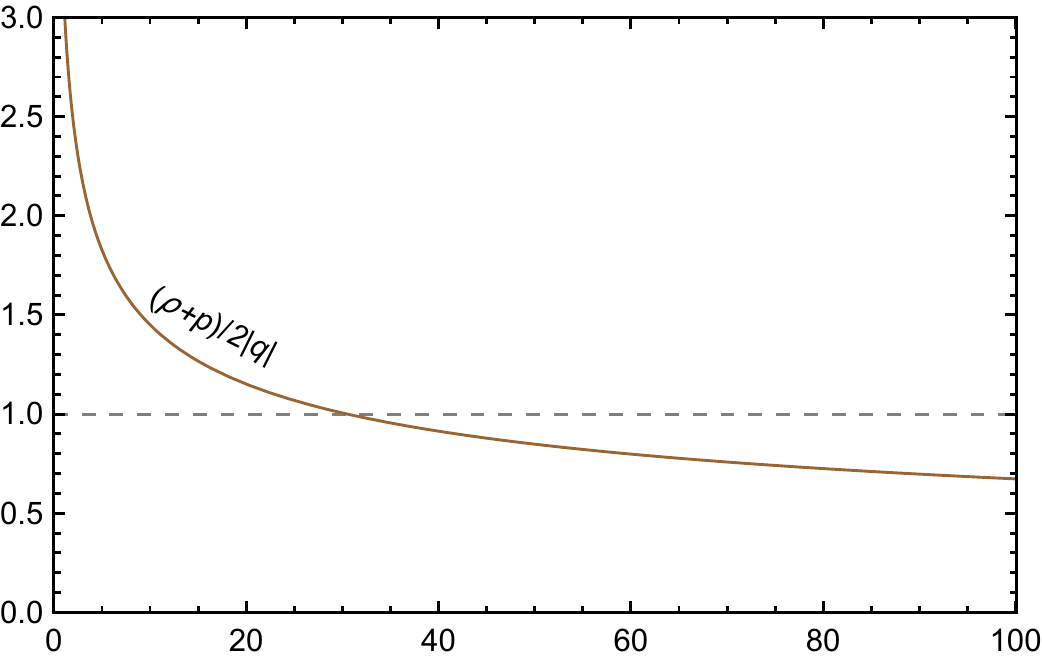}
 \caption{The curve for $\rho/2|q|$ in the case $a(t)=\left(t/t_0\right)^{2/3}$, 
with $t_0=1$ and $r=2.5\times m$ (we have chosen $m=1$).}
  \label{fig:pOrho-dust} 
\end{figure}

The energy density, the pressure, and the energy flux are given
by Eqs.~\eqref{prhodust}.
As in the case of the last section, the energy conditions are satisfied
at early times, but the heat flux decreases slower than 
the energy density and dominates for large times.
Such a behavior is depicted in Fig.~\ref{fig:pOrho-dust}.
In the present case Eq.~\eqref{eq:ratio} yields $n(t,r)=(\rho+p)/2|q|= 
{r^2}\Big/\left({m\,t_0^{2/3}t^{1/3}}\right)$. 
At very initial times (and fixed $r> 2m$),
the constraint $\rho+p \geq 2|q|$ holds and it can be verified that all the
energy conditions are satisfied. Moreover, since the
function $\rho +p$ vanishes faster than the energy flux $|q|$ with $t$,
at late times (and fixed $r>2m$) the constraint $\rho+p\geq 2|q|$
does not hold and the NEC, the WEC, and the SEC are violated.
On the other hand, since $|q|$ falls with $1/r^2$, for any fixed time
at sufficiently large radial coordinates ($r\rightarrow\infty$), 
the NEC, the WEC, and the SEC are satisfied.

Some details of the energy conditions along the trapping horizons and their
properties are given
in Appendix~\ref{app:horizonsclass}.

\subsubsection{\texorpdfstring{$a(t)=\left(\dfrac{t}{t_0}\right)^{1/3}$}{Lg}: 
Stiff matter model} \label{sec:stiff}

\begin{figure}[htb!]
\includegraphics[width=.4\textwidth]{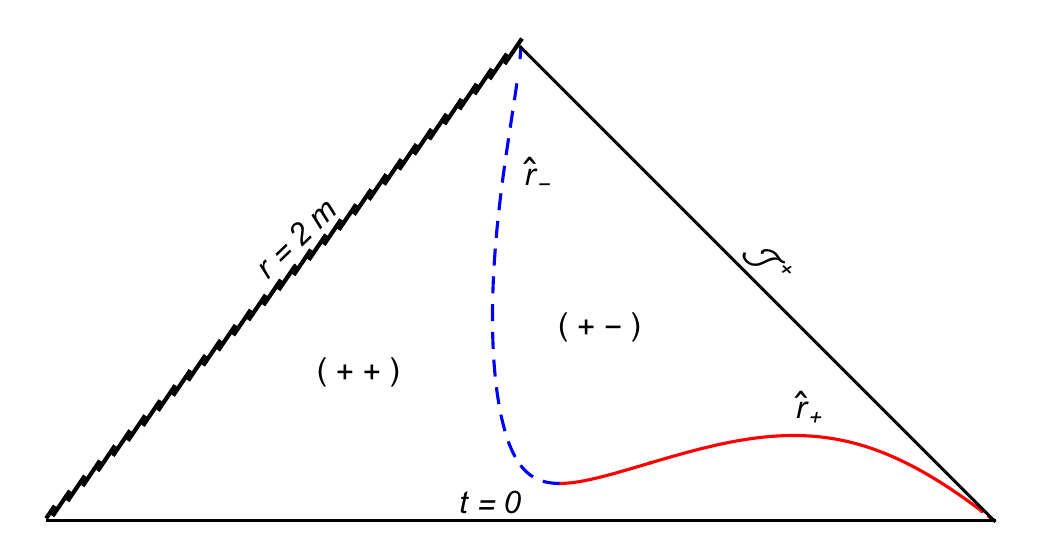}
 \caption{The causal diagram for the stiff matter model with $a(t) = 
\left(t/t_0\right)^{1/3}$, with $t_0=1=m$. The spacetime is 
geodesically incomplete with $r=2m$ and $t=0$ being singular surfaces. At 
late times, the inner 
trapping horizon is timelike while the outer horizon is spacelike.} 
  \label{fig:diagrama-stiff-matter}
\end{figure}

In some cosmological models there is a ``stiff matter'' era
\cite{Chavanis:2014lra}, notably when dark matter is modeled as Bose-Einstein 
condensates, where the equation of state is $\rho = p$. A stiff matter model 
with positive energy is considered with 
$a(t)=\left({t}/{t_0}\right)^{1/3}$, yielding an unbounded  
conformal time $\eta$. At late times the inner trapping horizon is timelike 
while the outer trapping horizons is spacelike for all times. The inner 
horizon changes to the spacelike 
character at a given intermediate time. The causal structure is presented in 
Fig.~\ref{fig:diagrama-stiff-matter}. As in the last two cases, the initial 
singularity $t=0$ is spacelike. Here, as in the case of dust matter, the 
surface $r=2m$ is singular and the spacetime is also future geodesically 
incomplete.

\begin{figure}[htb!]
 \includegraphics[width=.4\textwidth]{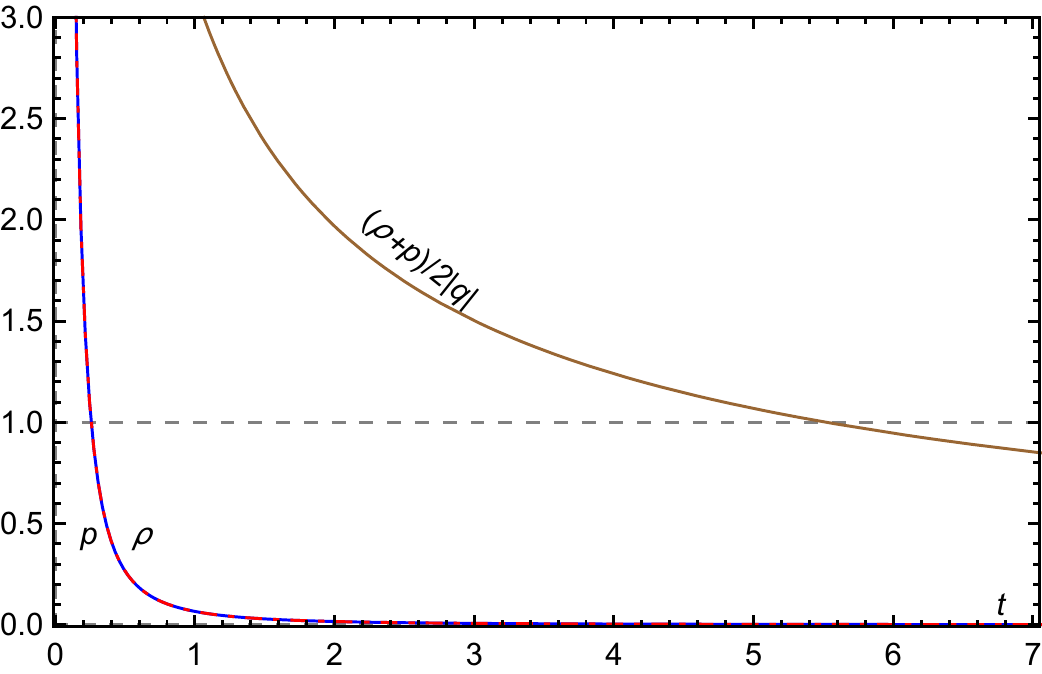}
 \caption{The energy density, the pressure, and the ratio $(\rho+p)/2|q|$ 
 from Eqs.~\eqref{prhostiff}, 
with $r=2.5m$ and $m=1$, for the case $a(t) = 
\left(t/t_0\right)^{1/3}$, with $t_0=1$. }
  \label{fig:prhostiff}
\end{figure}

The energy density, the pressure, and the energy flux for stiff matter model
are given respectively by
 \begin{equation} \label{prhostiff}
  8\pi\,  \rho =  8\pi\, p = \dfrac{1\ }{t^2\, f(r)},
  \quad |q| = \frac{2m}{3\,t_0^{1/3}t^{4/3}}\frac{1} {r^2f(r)},
 \end{equation}
where $f(r)$ is given by Eq.~\eqref{eq:f(r)}. 
The curves for $\rho$ and $p$ as a function of the cosmological time 
(for fixed $r$) are shown in Fig.~\ref{fig:prhostiff} 
(they are identical). The ratio 
$n(t,r)=(\rho+p)/2|q|={r^2}\Big/\left({2m\,t_0^{1/3}t^{2/3}}\right)$
is also shown in that figure.

As seen from the curves for $(\rho+p)/2|q|$ in
Figs.~\ref{fig:pOrho-dust} and \ref{fig:prhostiff},
the situation here is very similar to the case of dust matter reported in the
last subsection. We note that for early times the ratio $n(t,r)$ is grater
than unity, and we can show that the energy conditions are satisfied, 
while for late times the ratio becomes smaller than unity, so that the
condition $(\rho+p)-2|q|\geq0$ does not hold and the energy conditions
are violated. The curves shown in Fig.~\ref{fig:prhostiff} depend also
upon the radial coordinate $r$, so that for very large $r$ and finite $t$
the energy conditions are satisfied. For the behavior of the energy conditions
on the trapping horizons see Appendix~\ref{app:horizonsclass}.

\subsubsection{\texorpdfstring{$a(t)= \left(\dfrac{t}{t_0}\right)$}{Lg}: Fluid 
of cosmic strings model} \label{sec:linear}

A particularly intriguing case is when $a(t) =  \left({t}/{t_0}\right)$, 
where the mass function 
$M(t) = m a(t)$ increases linearly with the time $t$.

\begin{figure}[htb!]
\includegraphics[width=.4\textwidth]{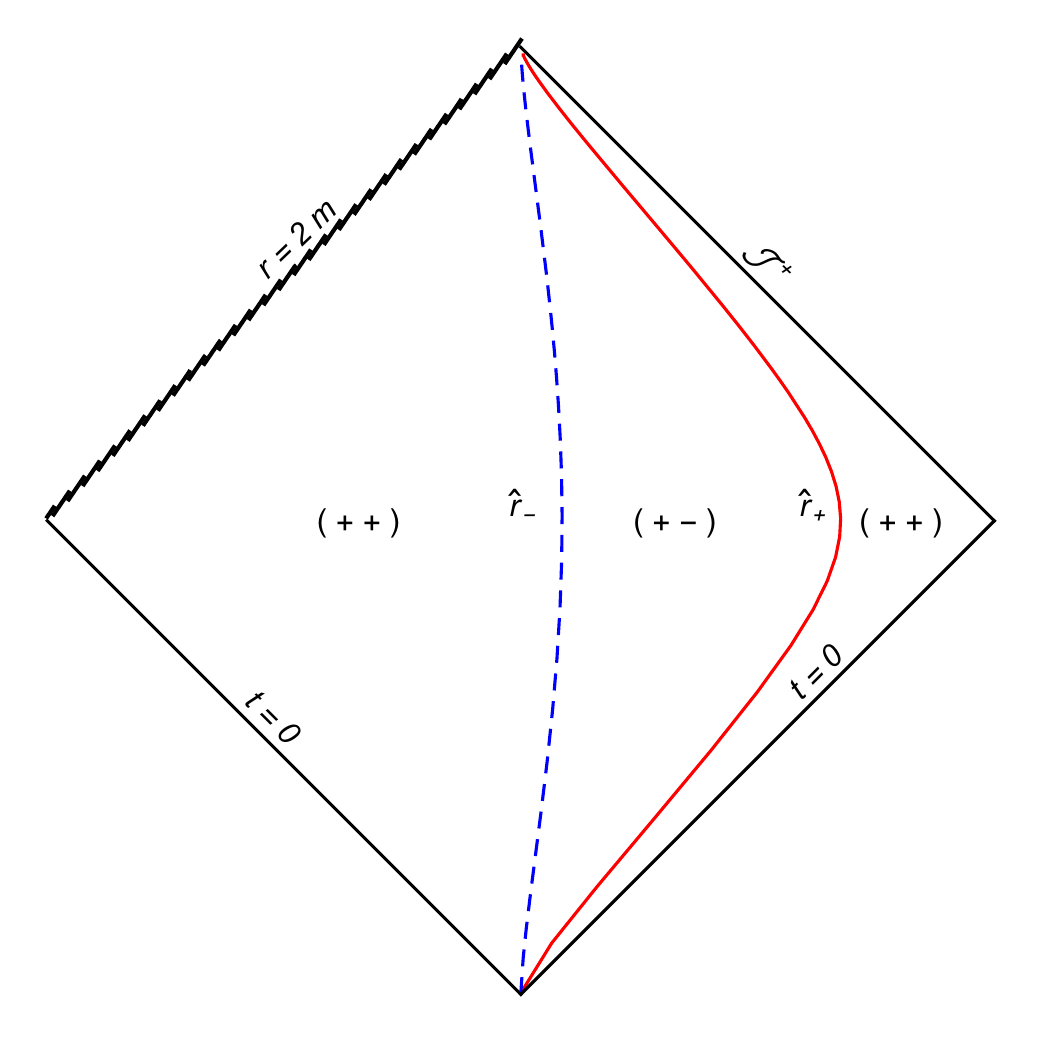}
 \caption{The causal diagram for $a(t) = \left(t/t_0\right)$. The initial 
singularity is 
lightlike. If $t_0 > 8 m$ the trapping horizons are as shown in this plot, 
being timelike everywhere. Here we have used $t_0=10\,m$, with $m=1$. In 
such a case, the locus $r=2m$ is singular}
 \label{fig:diagrama-linear}
\end{figure}

Taking $t_0 > 8m$ the 
trapping horizons are created with the big-bang at $t=0$ and persist for 
all 
times up to $t\to \infty$, being timelike for 
all times. The initial singularity is lightlike, and the locus $r=2m$ is 
singular, as indicated in the causal diagram depicted in 
Fig.~\ref{fig:diagrama-linear}. The spacetime is geodesically incomplete to 
the future, but all the incomplete geodesics terminate at the singularity 
$r=2m$.
 
The special case with $t_0 = 8m$ presents just one trapping horizon [the 
two solutions $\hat r_{\pm}(t)$ coincide], the regular region $(+ -)$ 
disappears but the conformal boundaries are the same as for $t_0> 8m$, as 
depicted in Fig.~\ref{fig:diagrama-linear}. On the other hand, for $t_0< 8m$ 
there are no trapping horizons and this situation is not considered in the 
present analysis. 

The scale factor $a(t)= (t/t_0)$ corresponds to the cosmological model of a 
universe filled by a fluid for which $p/\rho = -1/3$. 
Historically, the equation of state of such a form has been
related to a fluid of cosmic strings \cite{Letelier:1984dm,Vilenkin:1984rt}.  
However, with the advent of the 
accelerated expansion of the Universe, a plethora of dark energy models with 
$p/\rho <0$ have been formulated. In particular, successful scalar field 
models imply in a relation $\omega =p/\rho$ which varies with the 
cosmological time and assumes negative values at late times, including the 
$\omega =-1/3$ as a particular situation.

The energy density, the pressure, and the energy flux are, respectively,
 \begin{equation} \label{prholinear}
  8\pi\,  \rho = \frac{-8\pi\, p} {3}= \dfrac{3}{t^2\, f(r)}, 
  \quad 8\pi |q|= \frac{2\,m\, t_0}{t^2}\frac{1}{r^2f(r)},
 \end{equation}
 where $f(r)$ is given by Eq.~\eqref{eq:f(r)}.
 \begin{figure}[htb!]
 \includegraphics[width=.4\textwidth]{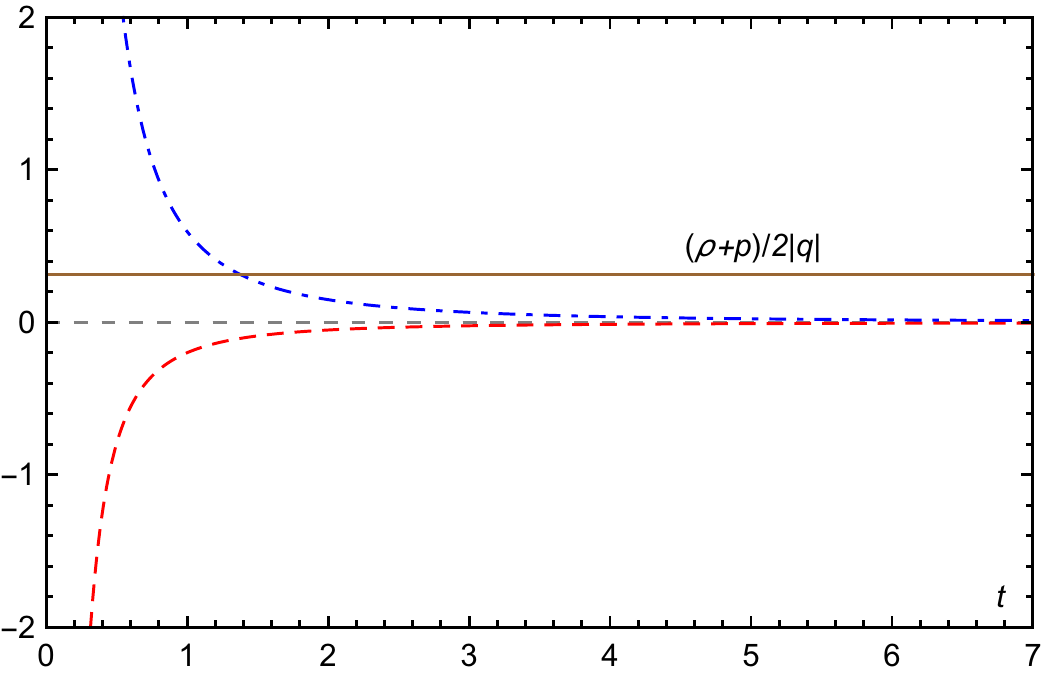}
 \caption{The energy density, the pressure, and the ratio $(\rho+p)/2|q|$, 
 from Eqs.~\eqref{prholinear} with $r=2.5m$, for the case $a(t) = 
t/t_0$, with $t_0=10m$ and $m=1$. }
  \label{fig:prholinear}
\end{figure}
The curves for $\rho$, $p$, and for the ratio $(\rho+p)/2|q|$
as a function of the cosmological time, for fixed $r=2.5m$,
are shown in Fig.~\ref{fig:prholinear}. 

Here one has $n(t,r)=(\rho+p)/2|q|=r^2/(2m\, t_0)$, which depends only on the radial
coordinate $r$, and is smaller than unity for small $r$. Since the
constraint $\rho + p \geq 2|q|$ is not fulfilled,  the 
energy conditions are violated at small scales. For $r\geq \sqrt{2m\,t_0}$ the NEC
and the WEC are satisfied, but the SEC is not. In fact, the SEC is violated 
everywhere in the spacetime.

 In the situation as in the case of Fig.~\ref{fig:diagrama-linear}, i.e., for $t_0> 8m$,
 the horizons are formed at $t=0$ with 
 $\hat r_\pm (t)=\left(1\pm\sqrt{1-8m/t_0}\right)/(2t_0)$ and last forever.
The NEC and the WEC are satisfied for all times along $\hat r_+(t)$, 
but all the energy conditions are violated along $\hat r_-(t).$
See Appendix~\ref{app:horizonsclass} for more details.

\subsection{Solutions with bounded scale factor}
\label{sec:bounded}

\subsubsection{Preliminaries}

On the basis of the statement of Ref.~\cite{Price:2005iv} that a bound system 
either completely follows the cosmological expansion or is completely 
insensitive to it after some transient, it is reasonable to assume that, 
initially, the mass of the central object  follows the 
expansion and grows with the cosmological scale factor $a(t)$, $M=m\, a(t)$, 
but that latter it stops to follow expansion and the mass increases in a 
different rate. These conditions justify the choice of a different function 
$a(t)$ 
to describe the mass function $M$, at least for late times. 

Moreover, as already mentioned, an aim of the present work is to extract 
black-hole solutions from the Thakurta metric \eqref{eq:thakurta}, and so the 
$r=2m$ locus needs to be a nonsingular surface. In order to achieve such an 
objective, besides converging to a constant for large times, the study 
presented in Sec.~\ref{sec:r=2m} elucidated that the scale factor $a(t)$ must 
display a sufficiently fast decaying derivative $\dot{a}(t)$.

Besides the physical interpretation of the matter under collapse, the 
important point here is that the choice of a bounded scale factor $a(t)$ bears 
the idea that the time evolution of the ``local'' object depart from the 
evolution of the Universe as a whole. In this instance, the function 
$M(t)=m\,a(t)$ describes the 
time evolution of the localized object independently of the expansion of the 
Universe which would be described by a different scale factor.

\subsubsection{\texorpdfstring{$a(t)=\tanh^{2/3}(t/t_0)$}{Lg}: Initial power 
law model with smooth transition to vacuum} \label{sec:tanh23}

  With the last comments in mind, we try the function $a(t) = \tanh^{2/3} 
\,(t/t_0) $. This model simulates a dust dominated initial era, with 
$a(t)\sim (t/t_0)^{2/3}$, and a final phase with constant $a(t)=1$.  

\begin{figure}[htb!]
\includegraphics[width=.4\textwidth]{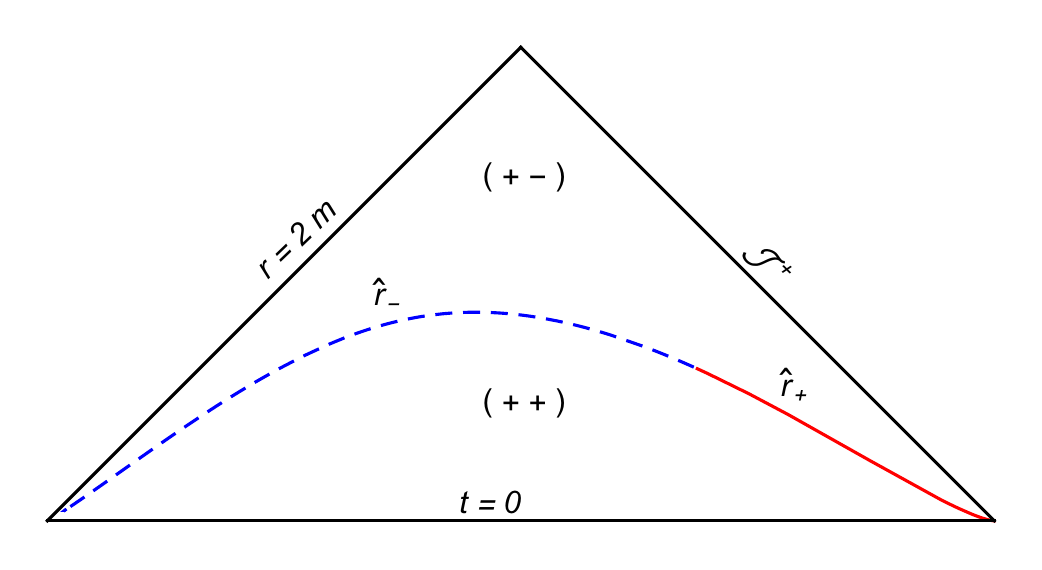}
 \caption{The causal diagram for $a(t) = \tanh^{2/3}\left(t/t_0\right) $, with 
$t_0=1=m$. The trapping 
horizons are spacelike everywhere, and the surface $r=2m$ is nonsingular. 
Also there is an initial spacelike singularity at $t=0$.}
 \label{fig:diagrama-tanh23}
\end{figure}

\begin{figure}[htb!]
\includegraphics[width=.48\textwidth]{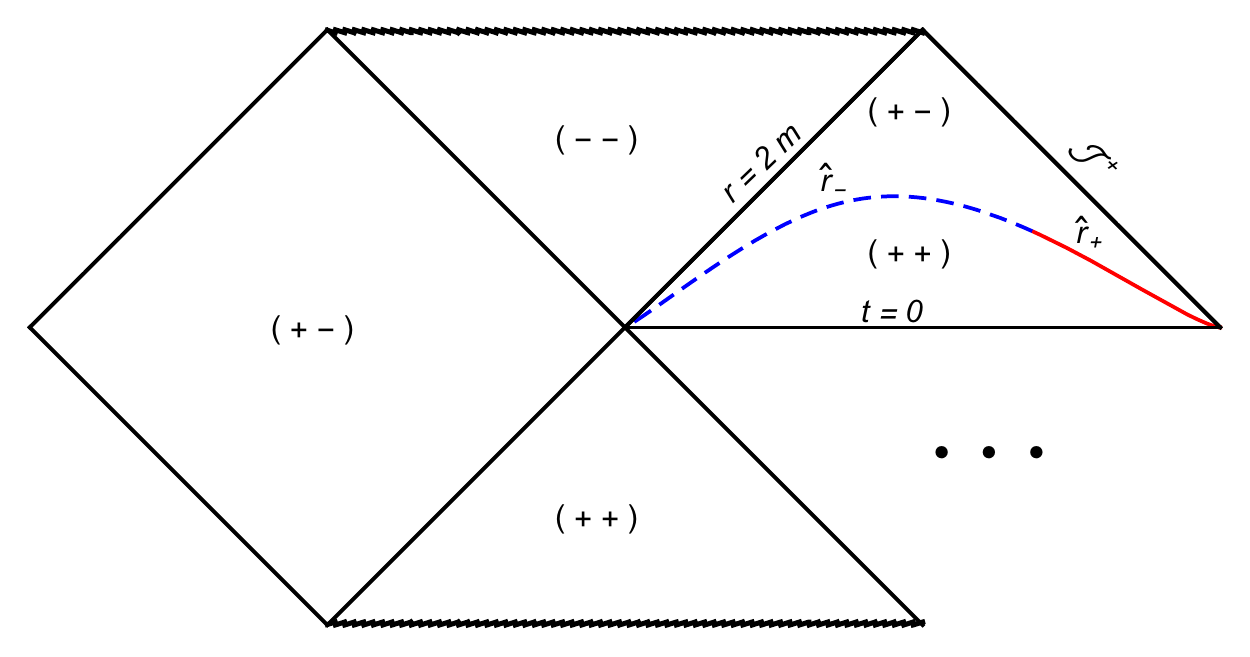}
 \caption {A possible extended causal diagram for $a(t) = 
\tanh^{2/3}\left(t/t_0\right) $, with $t_0=1=m$.
 The surface $r=2m$ is nonsingular and corresponds to a black-hole
horizon. The ellipses indicate that the diagram is to be continued.}  
 \label{fig:diagrama-tanh23ext}
\end{figure}

The scale factor $a(t)=\tanh^{2/3}(t/t_0)$ yields a spacetime whose incomplete 
causal diagram, for $t_0=1=m$, is given in Fig.~\ref{fig:diagrama-tanh23}. 
The surface $r=2m$ is nonsingular and the resulting spacetime is future 
geodesically incomplete along such a surface. 
Moreover, if we take $t_0 < 4m$ --- note that $t_0 = 2\tau$ --- then this 
spacetime can be extended and a possible extension is
represented in Fig.~\ref{fig:diagrama-tanh23ext}. Taking into account that in 
the limit $t \to \infty$ these Thakurta models converge to a Schwarszchild 
geometry, and that the trapping horizons are everywhere spacelike, then the 
gluing is made from a regular region. Therefore the extension is made by 
joining a trapped Schwarszchild region, followed by 
a regular Schwarszchild region, and so on. A more detailed and rigorous 
argument on this point is presented in Appendix~\ref{sec:B}, where we build 
Kruskal-like coordinates for the Thakurta spacetimes and discuss the 
conditions on its analytical continuation. 

 \begin{figure}[htb!]
 \includegraphics[width=.4\textwidth]{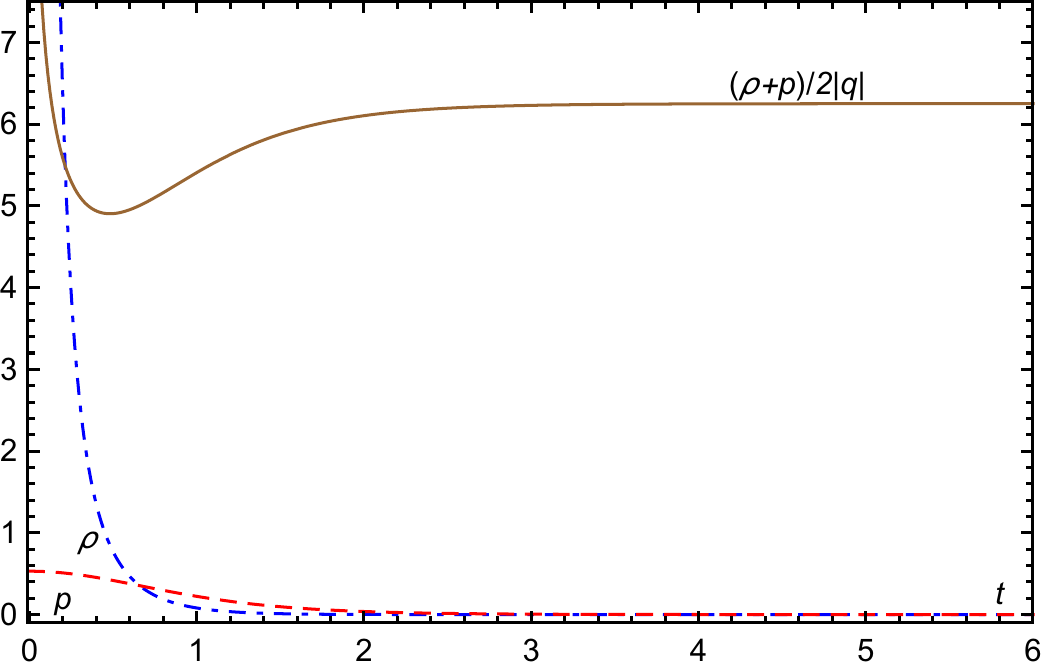}
 \caption{The energy density, the pressure, and the ratio $(\rho+p)/2|q|$,
 for the case $a(t) = \tanh^{2/3} 
\left(t/t_0\right)$, with $t_0=m$ and a fixed value of $r> 2m$ (here, 
$r=2.5m$ and $m=1$). }
  \label{fig:prhoTanh23}
\end{figure}

The background energy density, pressure, and energy flux in this case are,
respectively,  
\begin{equation}
 \label{prhoTanh23}
 \begin{split}
& 8\pi\,  \rho = \dfrac{16\,{\rm csch}^2\left(2t/t_0\right)}{3\,t_0^2 f(r)},
    \qquad 8\pi\, p 
    = \frac{8\,{\rm sech}^2\left(t/t_0\right)} {3\, t_0^2f(r)} ,\\
 &  8\pi\,|q| 
    =\frac{4m}{3r^2\, t_0} \frac{{\rm sech}^2\left(t/t_0\right)} 
    {\tanh^{5/3}\left(t/t_0\right) f(r)} ,    
 \end{split}
\end{equation} 
where $f(r)$ is given by Eq.~\eqref{eq:f(r)}.
 For small times it holds $a(t) \sim (t/t_0)^{2/3}$, and so the energy 
density 
approaches the cold dark matter cosmological model, $8\pi\,\rho\sim 
4\big/\left(3\, t^2f(r)\right)$ 
but the pressure is constant with time in that regime, 
$8\pi p\sim 8\big/\left(3\,t_0^2f(r)\right)$. For large times both the
energy density and pressure vanish, with the energy density going to zero
faster than the pressure. The ratio $p/\rho$ 
gives $2\,\sinh^2\left(t/t_0\right)$, which grows with $e^{2t/t_0}$ at large
times. This is shown in Fig.~\ref{fig:prhoTanh23}. 

To check the energy conditions let us then take the 
ratio $\left(\rho+p\right)/2|q|$ from Eq.~\eqref{eq:ratio},
which gives, $n(t,r)= \dfrac{r^2}{2m\, t_0}
\left(1 +\coth^2\left(t/t_0\right)\right)\tanh^{5/3}\left(t/t_0\right)$.
For small times it results  $ n(t,r) \sim r^2\Big/
\left(2m\, t_0^{2/3}t^{1/3}\right)$, while for large $t$
it tends to $n(t,r)\to r^2/\left(mt_0\right)$. Hence, at initial 
times and fixed $r<\infty$ the constraint $(\rho+p)\geq 2 |q|$ does not hold
and the energy conditions are not satisfied. 
On the other hand, for large times and with 
fixed $r^2>2mt_0$ the energy flux is always smaller than
the energy density and then the constraint $\rho+p\geq 2|q|$ is satisfied,
which assures that the NEC is satisfied. Moreover, owing to the fact that in
the present case one has $\rho+ p>0$, the WEC and the SEC are also satisfied
for large times.

The analysis of the energy conditions and properties of the trapping horizons are 
presented in Appendix~\ref{app:horizonsclass}.

\subsubsection{\texorpdfstring{$a(t)=\tanh(t/t_0)$}{Lg}: Cosmic strings type 
fluid  model with smooth transition to vacuum} \label{sec:tanh}

Let us now analyze a case with an initial singularity that is not 
spacelike and with bounded $a(t)$ at large times. This is the case for 
$a(t)=\tanh\left(t/t_0\right)$. At early 
times $t\to 0$ the scale factor goes as $a(t) = t/t_0$ and tends to unity for 
large times. The ratio between pressure and energy density for early times 
is 
$p/\rho =-1/3$, representing a fluid of cosmic strings 
\cite{Letelier:1984dm,Vilenkin:1984rt}.
This solution presents three possible structures depending on the value of 
$t_0$ that influence the behavior of the horizons. The analysis of this 
behavior is presented in Appendix~\ref{sec:asymptotic-horizon}.
 
\begin{figure}[htb!]
 \includegraphics[width=.48\textwidth]{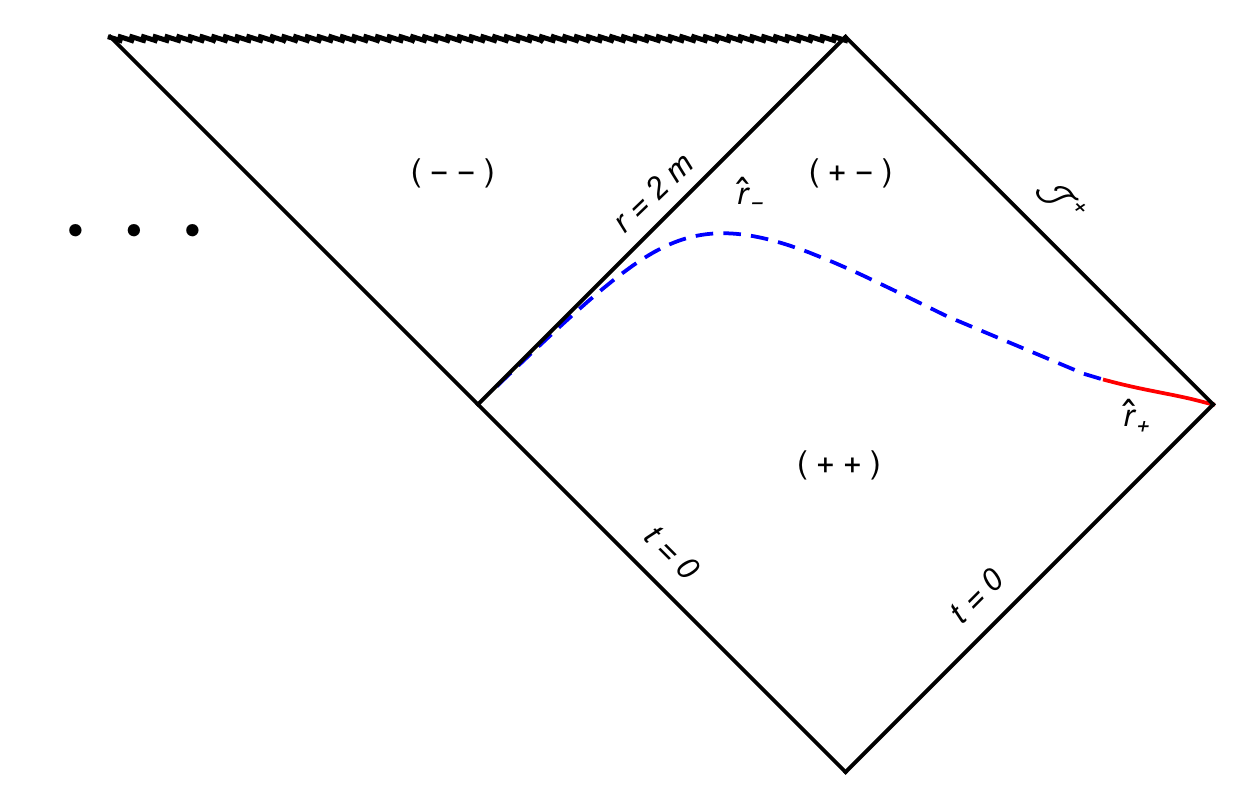}
 \caption{The extended causal diagram for the case $a(t) = 
\tanh\left(t/t_0\right)$, with $t_0=3m$ and $m=1$. The initial singularity 
is lightlike, and the surface $r=2m$ is nonsingular and corresponds to a 
black-hole horizon. The trapping horizons $\hat r_- $ and $\hat r_+$ are 
spacelike for all times.}
  \label{fig:diagrama-Tanh3}
\end{figure}

\begin{figure}[htb!]
 \includegraphics[width=.48\textwidth]{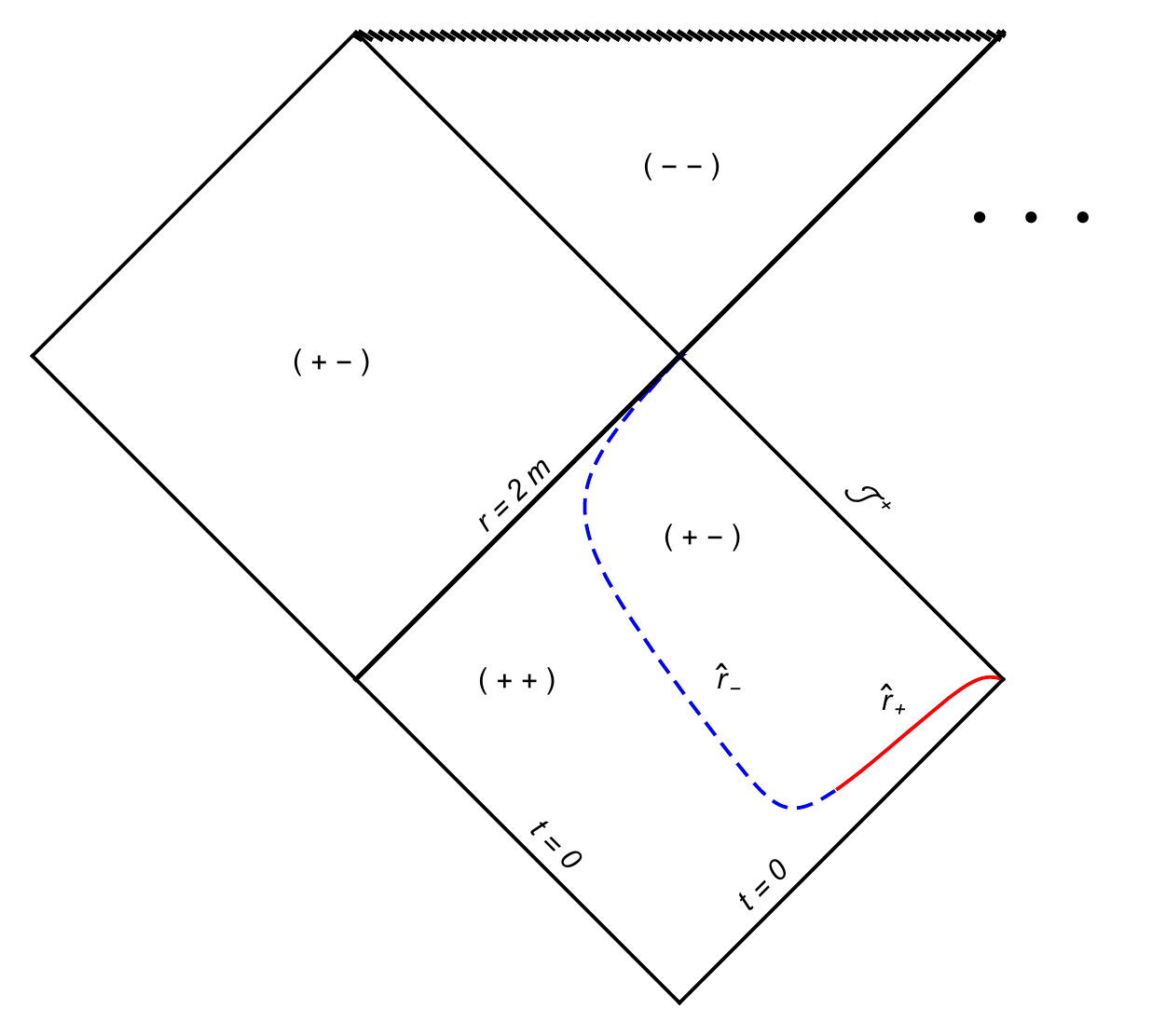}
 \caption{The extended causal diagram for the case $a(t) = 
\tanh\left(t/t_0\right)$, with $t_0=6m$ and $m=1$. The 
inner trapping horizon is timelike for late times, but it is spacelike in a 
small region of the spacetime at early times, close to the time the horizons 
are formed. The outer horizon is timelike for 
early times and becomes spacelike at late times. The initial singularity is 
lightlike, and the surface $r=2m$ is nonsingular corresponding to a white-hole 
horizon. }
  \label{fig:diagrama-Tanh6}
\end{figure}

\begin{figure}[htb!]
 \includegraphics[width=.4\textwidth]{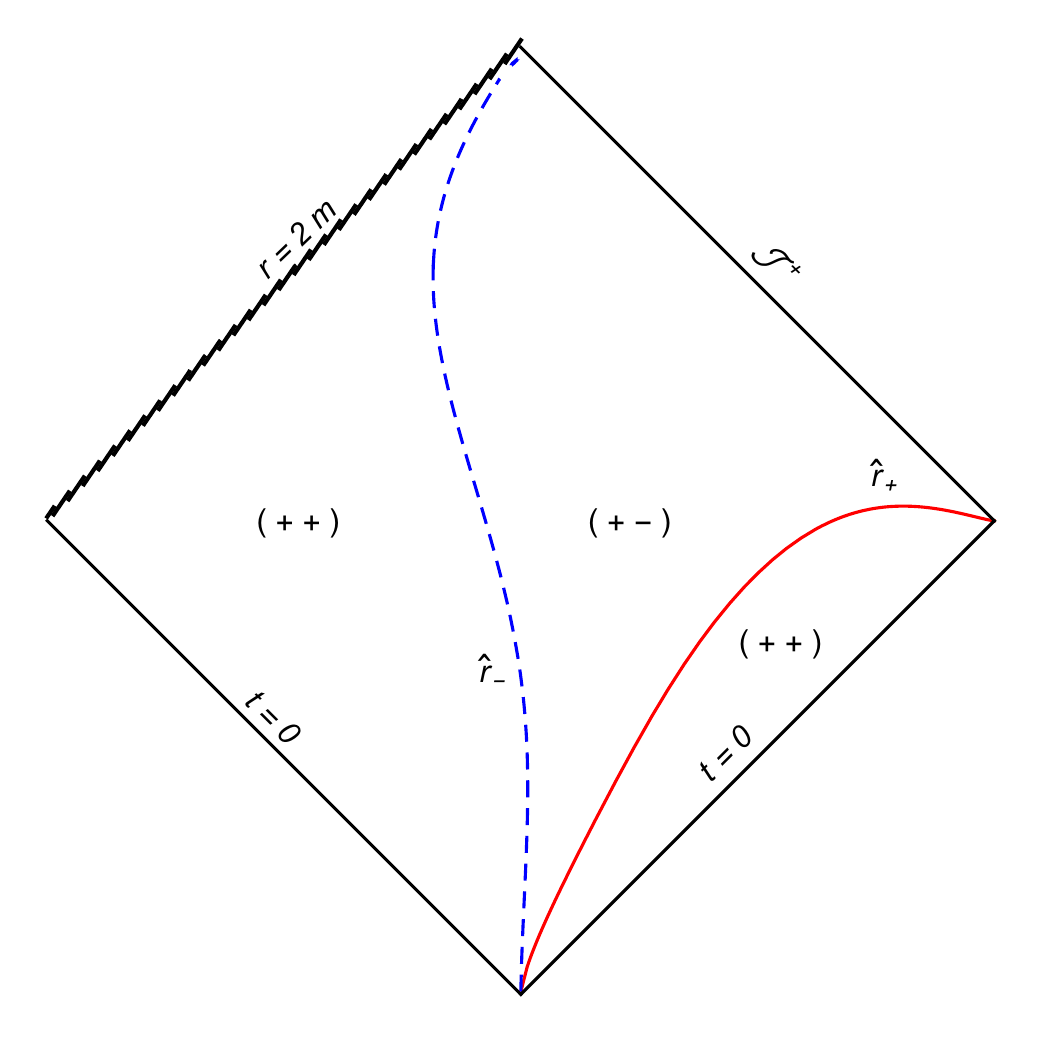}
 \caption{The causal diagram for the case $a(t) = \tanh\left(t/t_0\right)$, 
with $t_0=10m$ and $m=1$.
The inner horizon $\hat r_-$ is timelike, while the outer horizon $\hat r_+$ 
is timelike for early 
times and becomes spacelike at late times. The initial singularity is 
lightlike, and the surface $r=2m$ is singular. Notice that the horizons exist 
since the beginning of time.}
  \label{fig:diagrama-Tanh10}
\end{figure}

Considering $t_0<4m$ the horizons are spacelike at all times as showed in 
Fig.~\ref{fig:diagrama-Tanh3}. For $4m<t_0<8m$ the horizon $\hat r_-$ 
is timelike at large times
and the diagram is represented in Fig.~\ref{fig:diagrama-Tanh6}. Also 
considering $t_0>8m$, represented in Fig.~\ref{fig:diagrama-Tanh10}, the 
horizon $\hat r_-$ is timelike and both horizons exist at all times.

 In the case  $ a(t)= \tanh\left(t/t_0\right)$, the energy density, the
pressure, and the energy flux are given, respectively, by
\begin{equation}\label{prhoTanh}
 \begin{split}  
  & 8\pi\,  \rho = \dfrac{12\,{\rm csch}^2\left(2t/t_0\right)}{t_0^2\, f(r)},
    \\
 &   8\pi\, p     = \bigl[ 2 \cosh\left(2t/t_0\right)-3\bigr] \frac{4\, 
    { \rm csch}^2(2t/t_0)} {t_0^2\, f(r)} , \\
& 8\pi |q| = \frac{2m}{t_0}\frac{\rm csch^2\left(t/t_0\right)}{r^2f(r)}, 
 \end{split} 
 \end{equation}
where $f(r)$ is given by Eq.~\eqref{eq:f(r)}.
 \begin{figure}[htb!]
 \includegraphics[width=.4\textwidth]{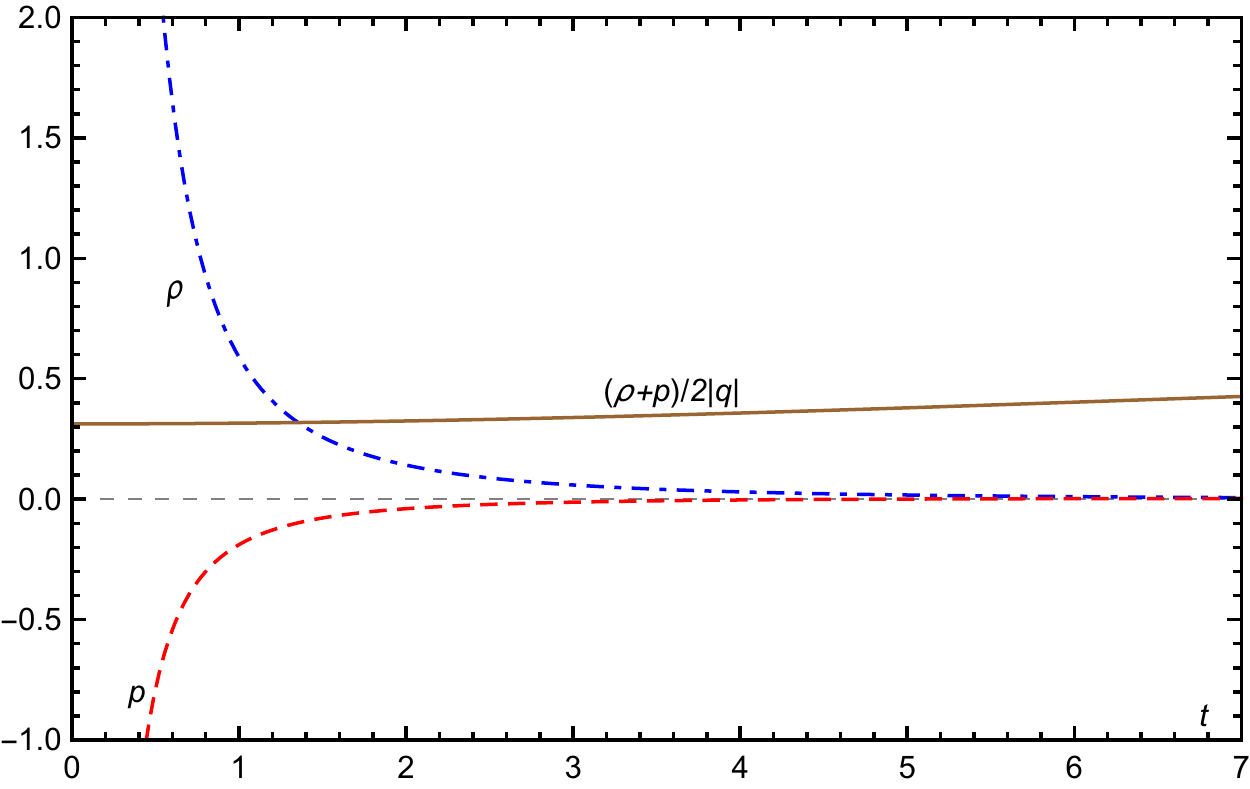}
 \caption{The energy density and the pressure from 
Eqs.~\eqref{prhoTanh} for the case 
$a(t) = \tanh\left(t/t_0\right)$,  with $t_0=10m$, $r=2.5m$ and $m=1$.}
  \label{fig:prhoTanh10}
\end{figure}
\begin{figure}[htb!]
 \includegraphics[width=.4\textwidth]{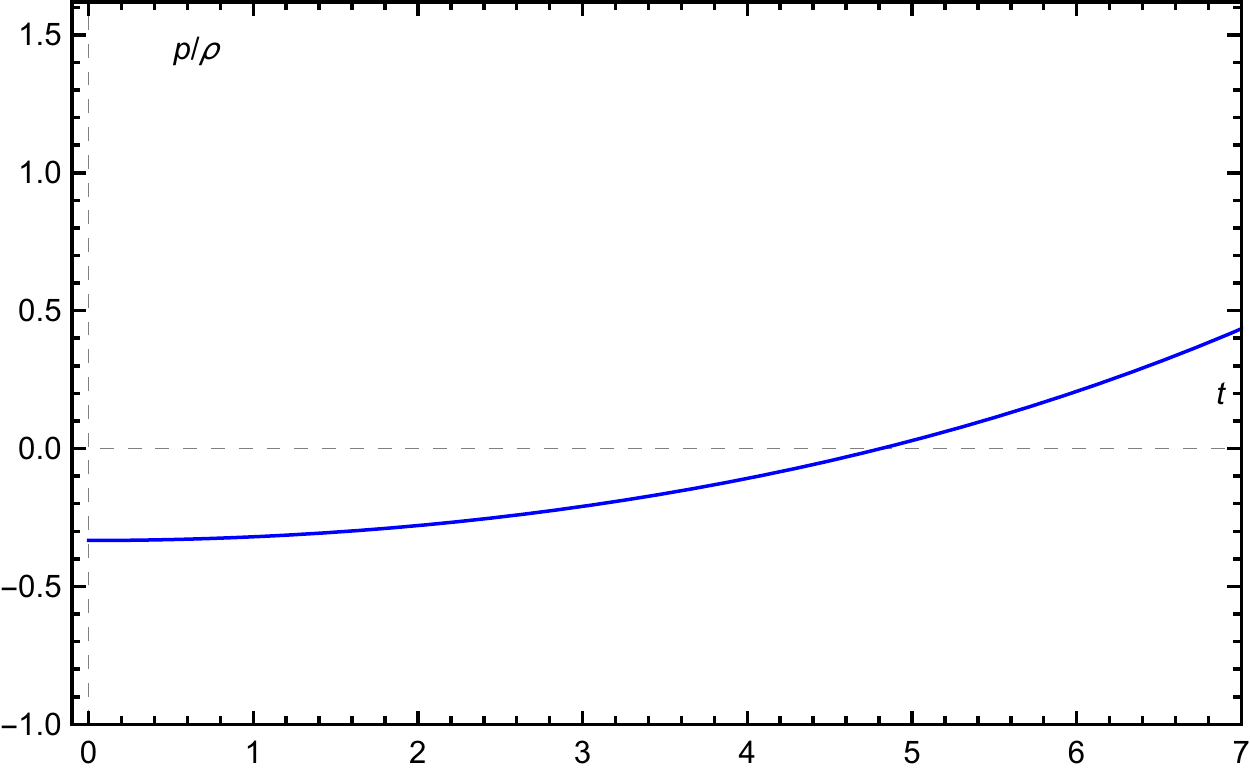}
 \caption{The ratio $p/\rho$ in the case $a(t) = \tanh\left(t/t_0\right)$,
 with $t_0=10m$ and $m=1$, as in Fig.~\ref{fig:prhoTanh10}. }
  \label{fig:pOrhoTanh10}
\end{figure}

As seen in Figs.~\ref{fig:prhoTanh10}~and~\ref{fig:pOrhoTanh10}, this scale 
factor gives a fluid whose energy density and pressure diverge at 
$t=0$ and vanish at $t\to\infty$. As in general for the Thakurta metric, the 
equation of state is such that $w=p/\rho$ depends on the cosmological time alone.
Here it is $w(t)= -1 +2\,\cosh\left(2t/2t_0\right)/3 $.
Function $w(t)$ assumes the value 
$p/\rho=-1/3$ at $t\to 0$, the equation of state for a fluid of 
cosmic strings \cite{Letelier:1984dm,Vilenkin:1984rt}.
The pressure grows to positive values and decays 
to zero in a rate slower than the energy density, and then the ratio 
$w(t)$ changes signs and assumes large values for large times, even though 
the fluid quantities are vanishingly small.

The ratio $(\rho+p)/2|q|$ starts with the value 
${r^2}/\left({2m\,t_0}\right)$ at $t=0$ and
reaches the value ${r^2}/\left({m\,t_0}\right)$ at $t\to\infty$. Hence, 
the (NEC, WEC and SEC) energy conditions are violated for sufficiently small
radial coordinate, but they are all satisfied for sufficiently large values
of $r$. 
This implies that the energy conditions tend to be violated along the 
$\hat r_-(t)$ branches of the trapping horizons, and are satisfied along 
the $\hat r_+(t)$ branch. In Appendix \ref{app:horizonsclass} we perform 
a more detailed analysis on such a subject.

\subsubsection{Truncated scale factor} \label{sec:truncated}

Some of the models with unbounded scale factor can easily be modified to 
describe a spacetime without a singularity at $r=2m$.
Considering that the accretion of matter/energy into the central object
eventually ceases because the whole matter of the surroundings has already 
been exhausted, we may take a different scale factor $a(t)$ after a 
given sufficiently long time $T$. 
Notice that a nonsingular surface $r=2m$ can be formed if we choose 
$\dot{a}(t)$ to vanish faster than $\exp[-t/2m]$.
The simplest choice is to take $a(t)= \,$constant for $t>T$.

For instance, in the case of dust matter studied in Sec.~\ref{sec:dust} we 
may take
\begin{equation} \label{eq:power2cutoff}
  a(t) = \left\{ 
  \begin{array}{cc}
               \left({t}/{t_0}\right)^{2/3}, & t\leq T,\\
               \left({T}/{t_0}\right)^{2/3}, & t> T,
                \end{array}\right.
\end{equation}
where $T$ is an arbitrary timescale. 
In this case, the mass function $M(t) = m a(t)$ initially grows by accretion 
of dust fluid (cold dark matter) at the same rate as the background expands, 
but suddenly stops growing at time $T$. This parameter $T$ is chosen as the
approximate cosmological time when the 
evolution of local object separates from the universal expansion. Such a time 
scale may be thought of as the time when the accretion of matter by the object 
becomes negligible. 

\begin{figure}[htb!]
\includegraphics[width=.47\textwidth]{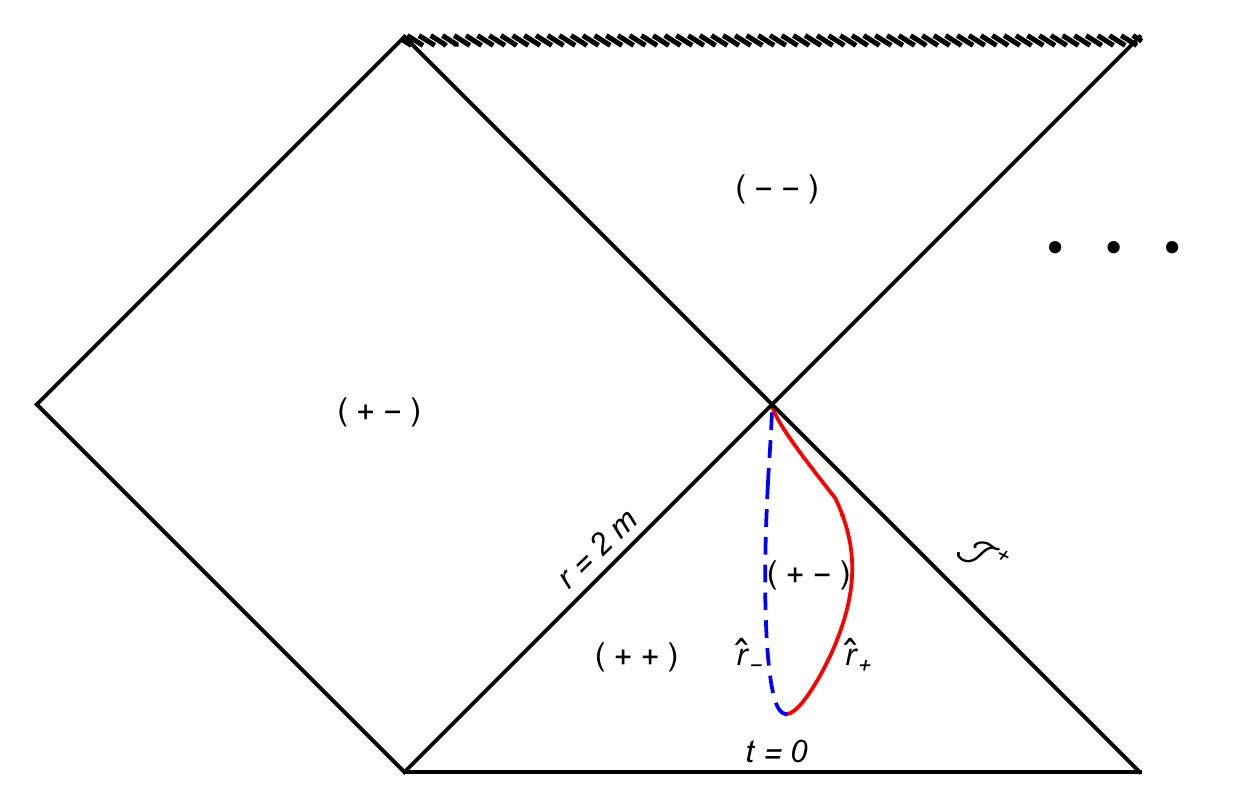}
 \caption{The extended causal diagram in the case the scale factor is 
truncated as $a(t) = \left(t/t_0\right)^{2/3} $ 
for $t\leq T$, and $a(t)=\left(T/t_0\right)^{2/3}$= constant for $t> T$. Here 
we have chosen $T=50m$, $t_0=m$, and $m=1$. The initial singularity 
$t=0$ is spacelike. 
The locus $r=2m$ is nonsingular and corresponds to a white-hole horizon.}
 \label{fig:diagrama-power2cutoff}
\end{figure}

The final spacetime is a white hole. In fact, the causal 
diagram is similar to what is presented in Fig.~\ref{fig:diagrama-dust}, 
but now the locus $r=2m$ is nonsingular, the resulting geometry is 
geodesically incomplete and the spacetime can be extended as shown in 
Fig.~\ref{fig:diagrama-power2cutoff}, where the extension is made by attaching 
a regular region, followed by a trapped region, into the antitrapped region.

\begin{figure}[htb!]
\includegraphics[width=.47\textwidth]{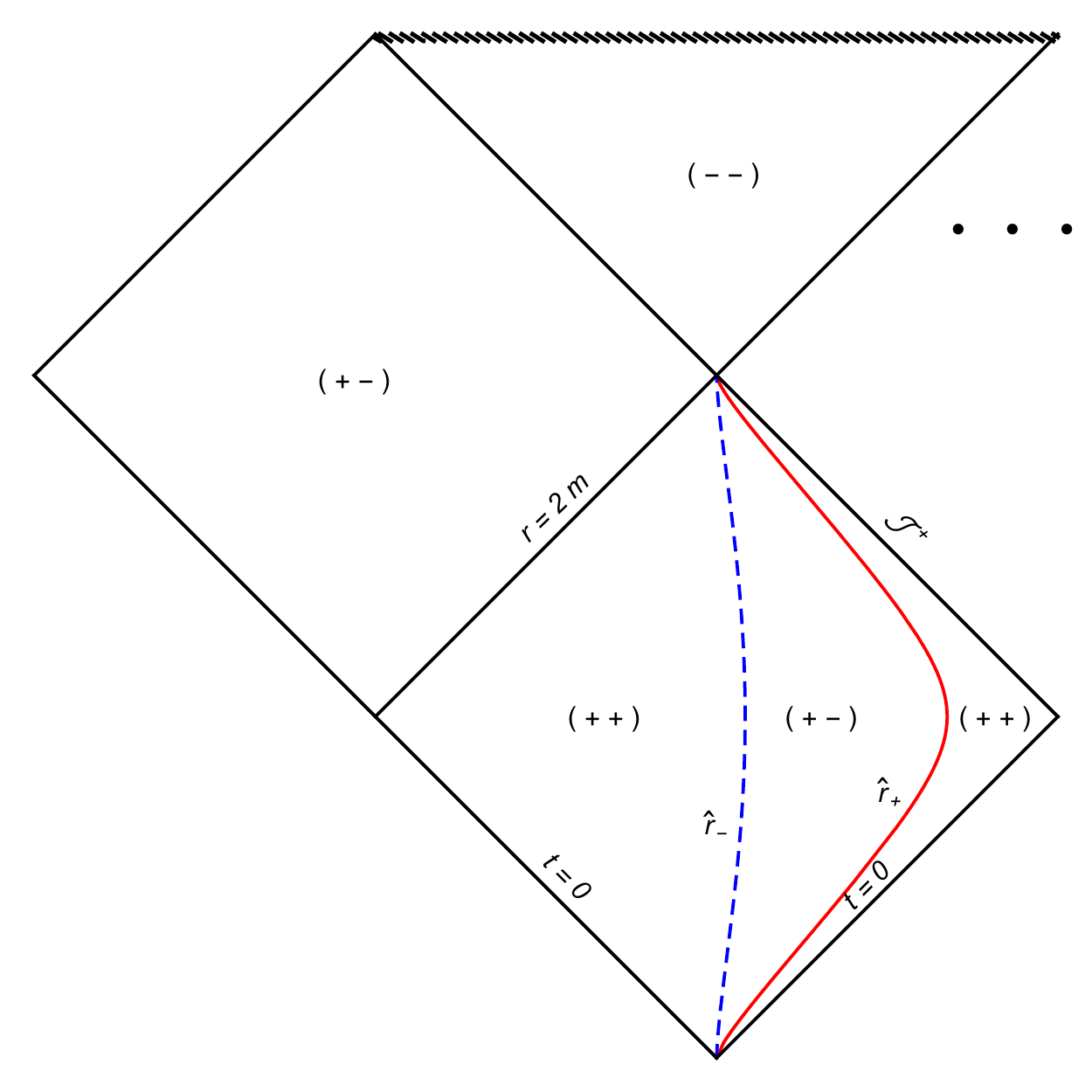}
 \caption{A possible extension in the case $a(t) = t/t_0$, for $T\leq T$, and 
$a(t) = T/t_0$, for $t> T$. Here we have chosen $T=t_0=10m$, and $m=1$. 
In such a case the surface $r=2m$ is nonsingular and corresponds to a 
white-hole horizon.}
 \label{fig:diagrama-linearext}
\end{figure}

A second interesting example where the scale factor may be truncated after a 
given time is the case considered in Sec.~\ref{sec:linear}. The scale factor 
is of the form
\begin{equation} \label{eq:strings2cutoff}
  a(t) = \left\{ 
  \begin{array}{cc}
              {t}/{t_0}, & t\leq T,\\       
             {T}/{t_0} , & t> T,
                \end{array}\right.
\end{equation}
where $T$ is an arbitrary timescale. The corresponding conformal diagram is 
shown in Fig.~\ref{fig:diagrama-linearext}.

\begin{figure}[htb!]
\includegraphics[width=.48\textwidth]{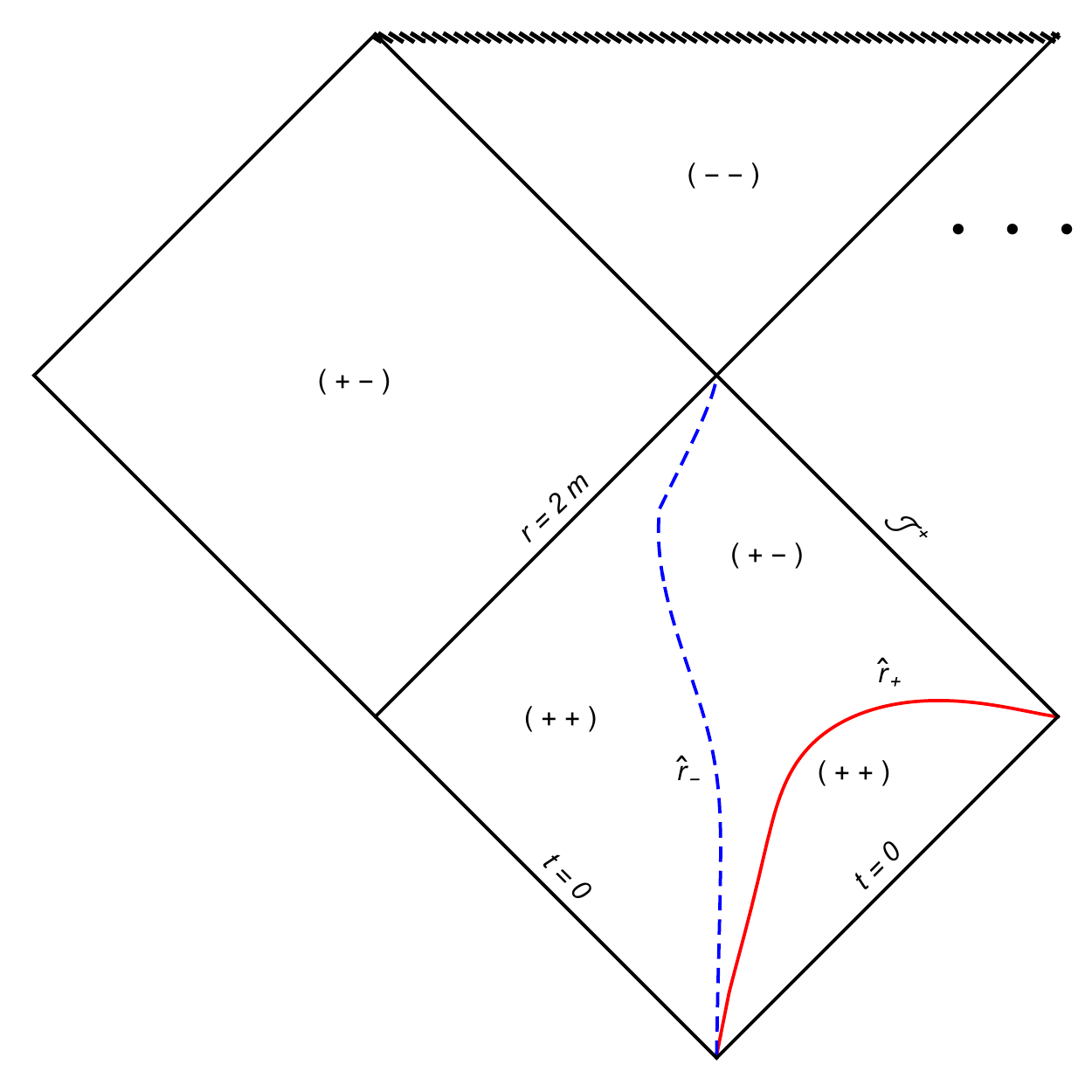}
 \caption{A possible extension for the case $a(t) = \tanh\left(t/t_0\right)$ 
for $t< T$ and $a(t) = \tanh\left(T/t_0\right)$ for $t \geq T$, with $T=50m$, 
$t_0=10m$, and $m=1$. The surface $r=2m$ is nonsingular and corresponds to 
a white-hole horizon.}
 \label{fig:diagrama-Tanh10ext}
\end{figure}

Is is also easy to obtain a black-hole type solution in the case of 
$a(t) = \tanh\left(t/t_0\right)$, with $t_0=10m$, studied above by truncating 
the scale factor in order to make it  tend to a constant faster enough.
Then, by choosing
\begin{equation} \label{eq:tanh10cutoff}
  a(t) = \left\{ 
  \begin{array}{cc}
              \tanh\left(t/t_0\right), & t\leq T,\\
               \tanh\left(T/t_0\right), & t> T,
                \end{array}\right.
\end{equation}
where $a(t)$ is a constant for $t> t_0$, an
extension of the spacetime represented by Fig.~\ref{fig:diagrama-Tanh10} can 
be found. 
With this, the locus $r=2m$, or $R= 2m a(t)$, is asymptotically a 
nonsingular 
lightlike surface.  The final state is a white hole.  
Figure~\ref{fig:diagrama-Tanh10ext} shows a possible analytical continuation 
of 
the spacetime corresponding to this truncated scale factor.

Let us stress that the models analyzed in this subsection should be
considered as first approximations to describe the {\it local} evolution
of the spacetime, with the term {\it  local} meaning the spacetime region
close to central object. In fact, the truncation process causes an abrupt
change in the spacetime evolution, and may introduce surface layers
which could spoil the physical properties of the spacetime. 
The alternative to avoid such layers is to replace the constant
$a$ for $t>T$ by a function $a_2(t)$ which falls off with $t$ 
in a rate such that $0<\dot a_2(t) < e^{-t/4m}$, 
and to try a smooth match at $t=T$.
Furthermore, the conditions $a(T)=a_2(T)$, $\dot a(T)= \dot a_2(T)$,
and $\ddot a(T)=\ddot a_2(T)$ assure the continuity of the energy density,
energy flux, and pressure at $t=T$. Then, a matching hypersurface of class
$C^2$ is enough to avoid introducing surface layers. Such a construction
does not change substantially the global causal structure of the spacetime, 
in the sense that the trapping horizons keep the asymptotic behavior and the 
surface $r=2m$ is regular at $t\rightarrow\infty$,
as in Figs.~\ref{fig:diagrama-power2cutoff}, 
\ref{fig:diagrama-linearext}, and \ref{fig:diagrama-Tanh10ext}, and 
leads to solutions with no shells/surface layers. The resulting $a_2(t)$
functions are given by cumbersome formulas and then we do not preset them 
here.

A possible issue remains in the case $a(t)=(t/t_0)^{2/3}$,
because in this case the dust character of the fluid is broken.  
The $C^2$ matching inevitably introduces a nonzero pressure for $t> T$, i.e.,
nevertheless continuous at $t=T$, the pressure is zero for $t\leq T$ but
it would not be zero for $t>T$ anymore.
In order to fix the pressure to zero also in the region $t>T$, 
a surface layer must be present at $t=T$, and so the curvature scalar
would acquire $\delta-$function terms.
Such a situation is of little interest for the present analysis.

\subsection{Further considerations}

Many other examples can be constructed for different
choices of $a(t)$, however they will present essentially the same causal 
structure. This is due to the fact that the causal structure is 
determined by the asymptotic behavior of the chosen scale factor $a(t)$ at two 
limits, namely, $ t \to 0$ and $t \to \infty$. Therefore, we can craft any of 
the possible causal structures showed here, and still other ones that combine 
the possible behaviors of the big-bang singularity, the asymptotic character 
of the trapping horizons and the properties of the $r=2m$ surface and its 
continuation beyond the horizon (if it is nonsingular) by building 
functions 
$a(t)$ with the required properties at those two limits.

\section{Conclusion} \label{sec:conclusion}

In this paper we performed a thorough study of the spacetimes built by a 
conformal transformation of the Schwarszchild metric in Schwarszchild 
coordinates, using a time-dependent scalar factor $a(t)$. We demonstrated that 
for unbounded scale factors $a(t)$, the resulting models do not describe black 
holes, but inhomogeneous expanding universes presenting a regular 
region with a finite lifespan or even spacetimes with a null singularity at a 
finite proper time in the future.

In order to obtain black-holes spacetimes, we had to 
resort for bounded scale factors, with a rapidly vanishing time derivative 
$\dot{a} \sim e^{-t/\tau}$. For instance, in the models with 
$a(t)=\tanh(t/t_0)$ the time derivative of the scale factor goes as $\dot{a} 
\sim e^{-2t/t_0}$. 
The properties of these models, even with similar functional forms, strongly 
depend on the value of $\tau$.  We have found that
\begin{enumerate}[(1)]
 \item For $\tau > 4m$, the spacetime has a future null singularity and is not 
extensible;
 \item For $2m < \tau < 4m$, the spacetime presents a white-hole horizon 
at $r=2m$;
 \item For $\tau < 2m$, the spacetime presents a black-hole event horizon at 
$r=2m$.
\end{enumerate}

The strong dependence of the causal structure on the asymptotic form of the 
scale function $a(t)$ is reminiscent of the cosmological black-holes solutions 
based on the McVittie metric, that can also represent white holes depending on 
the chosen parameters, as recently shown in 
Refs.~\cite{daSilva:2012nh,daSilva:2015mja}.  The important difference here is 
that black-hole models only suffer cosmological expansion during a finite time 
scale given by the parameter $\tau$, being essentially static for large times. 
This behavior may seem unphysical at first glance, but it can be thought of as 
a decoupling model of an initially bound system that, after a transient, 
decouples from the cosmological expansion, similar to the behavior of the 
classical atom studied in Ref.~\cite{Price:2005iv}. Therefore, this can be an 
interesting model for dynamical, accreting black holes, that after the 
accretion time decouples from the cosmological expansion and ceases to 
accrete, and that at late times lives in an effectively static and empty 
bubble. 

This analysis shows how subtle the interpretation of 
apparently simple solutions of general relativity may be and provides 
examples of black holes and white holes with time-dependent mass and dynamical 
horizons. These results indicate that this construction of dynamical solutions 
from static ones can be used for other classes of black-hole solutions, as 
Thakurta's original attempt using Kerr line element. It is important to remark 
that this method of construction using a time-dependent conformal factor is 
dependent on the choice of the coordinates in the original metric. For example, 
in the case of Schwarszchild metric, a different choice of coordinates leads to 
the Sultana-Dyer construction \cite{sultana}. 

Finally, we wonder that this method of building dynamical black holes from 
static ones may be employed to produce new solutions in a number of 
applications in black-hole physics, such as the AdS/CFT correspondence, 
black-hole thermodynamics and properties of general relativity in the strong 
field regime.

\begin{acknowledgments} 
We thank B. R. Majhi for pointing out a misleading statement in the first draft.
We also thank Coordena\c{c}\~ao de
Aperfei\c{c}oamento do Pessoal de N\'\i vel Superior (CAPES), Brazil, for
scholarships and Grant No.~88881.064999/2014-01.
V. T. Z.  thanks Funda\c c\~ao de
Amparo \`a Pesquisa do Estado de S\~ao Paulo (FAPESP), Grant No. 
2011/18729-1, and
Conselho Nacional de Desenvolvimento Cient\'\i fico e Tecnol\'ogico of 
Brazil (CNPq), Grant No. 308346/2015-7.
\end{acknowledgments}
\appendix

\section{Asymptotic behavior of the trapping horizons} 
\label{sec:asymptotic-horizon}

In order to determine the character of the horizons hypersurfaces we 
investigate the character of the normal vector $n_\mu=\partial_\mu 
\hat{r}_\pm$. We consider only cases where $\dot{a} \to 0$ as $t \to \infty$. 
In the cases where $\dot{a} \nrightarrow 0$ the horizons have a 
finite span in the $R\times t$ subspace and do not exist for sufficiently 
large times.  The case $\dot{a}$ tends to a nonzero constant  as $t \to 
\infty$ requires a separated study. The hypothesis  
$\dot{a} \to 0$ greatly simplifies the expressions for the 
horizons by means of series expansions in terms of $\dot a$. 
Equation~\eqref{appr}  
can be approximated by
\begin{gather}
 \hat{r}_+ = \dfrac{1}{\dot{a}} +\mathcal{O}(\dot{a})  \,,\\
 \hat{r}_- = 2m \left(1 + 2m\dot{a} \right) + \mathcal{O}(\dot{a}^2)\, . 
\end{gather}
Therefore, the normal vectors to the trapping horizons $\hat{r}_\pm$ are 
given respectively by
\begin{gather}
 n^{(+)} = \frac{\ddot{a}}{\dot{a}^2}\,\ud t + \ud r\, ,\\
 n^{(-)} = -4m^2 \ddot{a} \, \ud t + \ud r\, .
\end{gather}

Taking the (squared) norm of $n^{(-)}$, for the inner horizon $\hat{r}_-$, we 
get
\begin{gather}
|n^{(-)}|^2 =-\frac{(4m^2)^2 \ddot{a}^2}{f[\hat{r}_-(t)])}
 +a^{-2}f[\hat{r}_-(t)]  \nonumber\\
= -\frac{(4m^2)^2 \ddot{a}^2}{2m\dot{a}} + \frac{2m\dot{a}}{a^2} 
\,,\label{norm-nminus}
 \end{gather}
\noindent
where we used the approximation
\begin{equation}
f[\hat{r}_-(t)] \approx 2m\dot{a}\,.
\end{equation}

Similarly, for the outer horizon $\hat{r}_+$ it follows
\begin{gather}
  |n^{(+)}|^2 = -\frac{1}{1-2m \dot{a}}\frac{\ddot{a}^2}{\dot{a}^4} +
a^{-2}\left(1-2m\dot{a}\right) \nonumber \\
= (1 + \mathcal{O}(\dot{a}))\left(\frac{1}{a^2}
-\frac{\ddot{a}^2}{\dot{a}^4} \right), \label{norm-nplus}
\end{gather}
\noindent
where the approximation
\begin{equation}
f[\hat{r}_+(t)] \approx 1- 2m\dot{a}\,,
\end{equation}
\noindent
was used.

With Eqs.~\eqref{norm-nminus} and \eqref{norm-nplus} at hand, in order to 
determine the asymptotic character of the trapping horizons, we only need to 
determine the sign of the normal to the horizon surfaces for a given 
asymptotic behavior of the scale factor $a(t)$.

For scale factors of the type $a(t) \sim t^{\alpha}$, with $0< \alpha < 1$, 
one has ${\ddot{a}^2}\big/{\dot{a}^4} \sim 1$,  
${\ddot{a}^2}\big/{\dot{a}^2} \sim t^{-2}$, and ${1}/{a^2} \sim
t^{-2\alpha}$. Applying these results to Eq. \eqref{norm-nminus} we obtain 
$|n^{(-)}|^2>0$ for large $t$. Therefore, 
\begin{enumerate}[(1)]
 \item 
$\hat{r}_-$ is timelike  for large $t$ if $0< \alpha < 1$.
\end{enumerate}

On the other hand, still for power law scale factors with $a(t) \sim 
t^{\alpha}$, the nature of the outer horizon depends upon $\alpha$. 
From Eq. \eqref{norm-nplus} we obtain that
\begin{enumerate}[(1)]
 \item  $\hat{r}_+$ is timelike for large $t$ if $\alpha > 1/2$,
 \item $\hat{r}_+$ is spacelike for large $t$ if $\alpha < 1/2$.
\end{enumerate}

In the case of models with a nonsingular surface at $r=2m$,  typically, the 
scale factor is such that $\dot{a} \sim e^{-t/\tau}$. With no loss of 
generality, we can set $a_\infty = 1$, since we can always achieve that by a 
rescaling $r \to { r}/{a_\infty}$, $m \to {m}/{a_\infty}$. Using 
Eqs. \eqref{norm-nminus} and \eqref{norm-nplus}, it can be shown that for 
large times $\hat{r}_-$ and $\hat{r}_+$ behave as follows,
\begin{enumerate}[(1)]
 \item $\hat{r}_-$ is spacelike if $\tau < {2m}$,
 \item $\hat{r}_-$ is timelike if $\tau > {2m}$,
  \item $\hat{r}_+$ is always spacelike. 
\end{enumerate}

\section{The nature of the horizon at \texorpdfstring{$\boldsymbol{r = 
2m}$}{Lg}} \label{sec:B} 

As seen above there are cases where the surface at $r=2m$ is a null horizon 
and the metric at that locus has the Schwarzschild form. In order to 
build an analytical continuation, we have to determine the kind of horizon 
it is. 
Since the metric is conformal to Schwarzschild, we can use Kruskal-like 
coordinates in order to extend it. Let $U$ and $V$ be such coordinates, 
defined by
\begin{gather}
U= \exp\left[\frac{\eta + r^*}{2m}\right]\,, \nonumber\\
V= - \exp\left[\frac{-\eta + r^*}{2m}\right]\,,
\end{gather}
\noindent
so that the metric \eqref{eq:conformal} can be cast in the form
\begin{gather}
\ud s^2 = a^2(\eta) \frac{e^{-r/2m}}{r} \left( -\ud U \ud V + \dots \right) 
\,,
\end{gather}
\noindent
which is well behaved at $r=2m$, that is given 
by $V=0$ in $(U,\,V)$ coordinates. The null expansions at $r=2m$ are 
\begin{gather}
 \Theta_+|_{r=2m} = V\,, \nonumber\\
 \Theta_-|_{r=2m} = U\,.
\end{gather}

Note that the null surface $r=2m$ is also a trapping horizon since $\Theta_+ 
=0$ there. Since $\Theta_+>0$ in the region covered by 
$(t,r)$ coordinates with $r> 2m$, by continuity it follows that  $\Theta_+ 
< 0$ for $r<2m$. This implies that 
the surface $r=2m$ of the Thakurta model should be identified to the $U$ axis 
of the half of a Schwarszchild spacetime defined by $V< 0$. 
 
In order to determine to which quadrant of the Schwarzschild maximal extension 
we can continue the Thakurta nonsingular models, we have to analyze the 
sign 
of the expansion of ingoing rays as they reach the horizon. Using $(t,r)$ 
coordinates again, we obtain 
\begin{gather}
 \Theta_- = \frac{2}{ar} k_{(-)}^\mu \nabla_\mu (ar) = 
\frac{2}{r}\left[\dot{a}r - f(r) \right]\,, 
\end{gather}
\noindent
which, evaluated for $r= r_-(t)$ and large $t$, gives
\begin{gather}
 \Theta_- \approx \frac{1}{m}\left(2m \dot{a} - e^{-t/2m}\right)\,.  
\label{thetalimit}
\end{gather}
Hence, assuming $\dot{a} \sim e^{-t/\tau}$, Eq.~\eqref{thetalimit} implies in
\begin{enumerate}[(1)]
 \item $\Theta_- < 0$ if $ \tau < 2m$,
 \item $\Theta_- > 0$ if $ \tau > 2m$,
\end{enumerate}
\noindent
which are the same conditions found for the asymptotic character of the inner 
trapping horizon $\hat{r}_-$ in Appendix~\ref{sec:asymptotic-horizon}. 

In the case where $\Theta_-< 0$ as $r_- \to 2m$, the horizon at 
$r=2m$ is the boundary of a regular region, and hence the analytical 
continuation must be into a trapped region of the Schwarzschild spacetime, 
corresponding to the region II ($V< 0$, $U<0$) of its maximal extension. 

In the case where $\Theta_-> 0$ as $r_- \to 2m$, the horizon at 
$r=2m$ is the boundary of an antitrapped region, and hence the analytical 
continuation must be into a regular region of the Schwarszchild spacetime, 
corresponding to the region I ($V<0$, $U>0$) of its maximal extension.

This argument justifies the analytical continuations of Thakurta spacetimes 
presented in Figs.~ \ref{fig:diagrama-tanh23ext}, \ref{fig:diagrama-Tanh3}, 
\ref{fig:diagrama-Tanh6}, \ref{fig:diagrama-power2cutoff},
\ref{fig:diagrama-linearext}, and 
\ref{fig:diagrama-Tanh10ext}.

\section{Geodesic completeness considering subleading terms} 
\label{app:subleading}

Here we extend the analysis of the geodesic completeness made in
Sec.~\ref{sec:geo-complete} including the contribution of
subleading terms.

We start defining $\Delta(t) = r_- - 2m$ so that 
Eq.~\eqref{dtdlambda} can be written in terms of $\Delta(t)$ as
\begin{gather}
 t'' = \left[ \frac{1}{2ma} \left(1 +\frac{r_- +
2m}{r_-^2}\Delta(t)\right) - \frac{\dot{a}}{a} \right] t'^2,
\end{gather}
\noindent
which gives, instead of Eq.~\eqref{eq:lambda2},
\begin{equation}\begin{aligned}
 \ln t' =   \frac{\eta}{2m} - \ln a + 
\frac{1}{2m}\!\mathlarger{\int^t} \frac{r_-(s) + 2m}{a(s)
r_-(s)}\Delta(s)\, \ud s  + c_1,
                \end{aligned} \label{eq:c2}
\end{equation}
where $c_1$ is an arbitrary integration constant. Let us stress that
Eq.~\eqref{eq:c2} is an exact relation.
Therefore, instead of Eq.~\eqref{eq:lambda3}, we find for the affine
parameter
\begin{gather}
 \Delta \lambda = c_2 \mathlarger{ \int_{t_0}^t}
\exp\left[{-\frac{\eta(u)}{2m}}\right] a(u)\,
 e^{ -J(u)} \, \ud u \label{eq:lambda-exact}
\end{gather}
where $c_2$ is another integration constant and we defined
\begin{gather}
 J(u)= \frac{1}{2m}\! \mathlarger{\int_{t_0}^u}\frac{\left(r_-(s) + 2m\right)}
 {a(s)r_-(s)}\Delta(s) \ud s.\label{eq:c4}
\end{gather}
The crucial difference between Eq.~\eqref{eq:lambda-exact} 
and Eq.~\eqref{eq:lambda3} is the presence of the second exponential term in
the integrand, $e^{-J(u)}$. The first exponential term, 
$e^{-{\eta(u)}/{2m}}$, is essentially the same as in 
Eq.~\eqref{eq:lambda3} (taking $K = 1/2m$ they become identical)
and then, to complete the analysis including subleading terms, 
just the behavior of the second exponential term as $t \to \infty$ must be
studied here. 

We first notice that, since $r_-(t) > 2m\,, \, \forall t>0$, it
follows that $\Delta(t)\geq 0$ and the ratio $(r_-+2m)/r_-$ is bounded, more
precisely one has $1 < \left(r_- + 2m\right)/{r_-}< 2\,,\, \forall t > 0$.
Therefore, it suffices to study the properties of the simpler expression
\begin{gather}\label{eq:I(u)}
 I(u) = \exp \left[\frac{1}{2m} \int_{t_0}^u \frac{\Delta(s)}{a(s)}
 \ud s \right] \,.
\end{gather}
Now it is convenient to consider separately the two possibilities,
\begin{enumerate}[(1)]
 \item $\lim_{t \to \infty} \Delta(t) > 0$, which corresponds to $\eta$
bounded from above.
 \item $\lim_{t \to \infty} \Delta(t) = 0$, which corresponds
 to $\eta$ unbounded from above.
\end{enumerate}

In the first case, if $\Delta$ tends to a finite positive value, then,
for any $s$ in the time interval considered $(t_0, u)$, 
there exist constants 
$\Delta_{b}\,,\, \Delta_{a}$ such that $0< \Delta_{b} \leq \Delta(s) \leq
\Delta_{a}$, implying in
\begin{gather}
\left[\eta(u) - \eta(t_0) \right]  \Delta_{b}  \leq 
 \mathlarger{\int_{t_0}^u} \frac{\Delta(s)}{a(s)} \ud s \leq   \left[\eta(u) -
\eta(t_0) \right]\Delta_{a} .
\end{gather}
Taking the limit $u \to \infty$ and applying to $I(u)$ we finally obtain,
\begin{gather}
 0< \exp\left[ - \frac{\Delta_{a}}{2m} \Big(\eta_{\infty} - 
 \eta(t_0) \Big) \right] < \lim_{u \to \infty} I(u) < \nonumber\\
\exp\left[ - \frac{\Delta_{b}}{2m} \Big(\eta_{\infty} - \eta(t_0)
\Big)\right]\, ,
\end{gather}
\noindent
which demonstrates that $I(u)$ is bounded by two strictly  positive
constant values, and so the subleading terms give no relevant contribution
to the convergence properties of $\Delta \lambda$. 
This fact guarantees that the study of Eq.~\eqref{eq:lambda3}
as the criterion for convergence performed
in Sec.~\ref{sec:geo-complete} is a strong result.

Now we examine the case where $\Delta(t) \to 0$ as $t \to \infty$. As in
Sec.~\ref{sec:geo-complete} we take a linear scale factor
$a(t) = a_0 t$, $a_0$ being a constant parameter.  
In this case we have to review the argument since there is
no $\Delta_{b}> 0$ that bounds $\Delta(t)$ from below. However,
Eq.~\eqref{r-solution} can be expressed in terms of $\Delta$
to find
\begin{gather}
 \exp\left[{1+\frac{\Delta(t)}{2m}}\right] \Delta(t) = 
 \left(\frac{t}{t_0}\right)^{-{1}/{2ma_0}}\,.
\end{gather}
\noindent
Replacing the last result into Eq.~\eqref{eq:I(u)} yields
\begin{gather}
 I(u) = \exp \left[ \frac{1}{2m} 
\mathlarger{\int_{t_0}^u}
\frac{t_0^{1/2ma_0}}{s^{1+1/2ma_0}} \exp\left({1 + 
\frac{\Delta(s)}{2m}}\right) \ud s \right] \,,
\end{gather}
\noindent
which is convergent in the limit $u \to \infty$. Therefore, 
taking into account the bounded character of $I(u)$, $0 < \lim_{u \to
\infty} I(u) < c_3$ for some constant $c_3 > 0$, the convergence of $\Delta
\lambda$ is determined by the first exponential term $\exp[{-\eta/2m}]$ as in 
 
Eq.~\eqref{eq:lambda3}, recovering
the results obtained in Sec.~\ref{sec:geo-complete}.

\section{The energy conditions and the nature of the trapping horizons}
\label{app:horizonsclass}

Here we investigate in some detail the character of the horizons within the
models presented in Sec.~\ref{sec:causal}.

A trapping horizon $h$ is defined from the condition $\Theta_-\big|_h=0$.
Correspondingly (following Ref.~\cite{Hayward:1993mw}), 
$h$ is said to be a past (future) trapping horizon
if $\Theta_+\big|_h>0$ ($\Theta_+\big|_h<0$). 
The horizon is outer (inner) if
${\cal L}_+\Theta_-\big|_h <0$ (${\cal L}_+\Theta_-\big|_h >0$), 
where ${\cal L}_+$ stands for the Lie derivative along the 
outgoing null geodesic lines. 

 The present analysis is based on the studies 
of Refs.~\cite{Hayward:1993mw,Andersson:2005gq}, which present theorems
that relate the chronological character of the trapping horizons to the
null energy condition. Here, for the sake of simplicity, we stick
to the statement given in Ref.~\cite{Hayward:1993mw}, though it is
weaker in general than the one presented in \cite{Andersson:2005gq}.
In the present models, however, since metric and horizons are spherically
symmetric, they are equivalent.  The statement is:

{\it Theorem I}: 
 \label{theorem}
If the null energy condition holds,  then an outer (inner)
trapping horizon is spacelike (timelike), and a trapping horizon is null
if and only if, additionally, the internal shear and normal energy density 
(energy flux) vanish.

\subsection{The case of Sec. \texorpdfstring{ \ref{sec:LCDM}}{Lg} }

In the situation of 
Fig.~\ref{fig:diagrama-Sinh23}, the horizons are formed at $t=t_i\simeq 0.31$,
with values $\hat r_\pm(t_i)=4.0$ and disappear at $t=t_f\simeq 62.82$
also with values $\hat r_\pm(t_i)=4.0$.
We have calculated the ratio  $n(t,r)$ [see Eq.~\eqref{eq:ratio}]
for $r=\hat r_\pm(t)$ and checked
the energy conditions along both horizons. 

\begin{enumerate}[(1)]
 \item Along $\hat r_+(t)$:  The NEC, the WEC and the SEC are satisfied since 
it is formed, but all the energy conditions are violated along $\hat r_+ (t) $ 
for times close to $t_f$. In particular, the NEC is satisfied up to $t_c\simeq 
38.73$. Close to the initial time $t_i\simeq 0.31$ one has  ${\cal 
L}_+\Theta_-<0$, but it becomes positive for times greater than $t_1\simeq 
0.44$. Hence, given that $\Theta_+ > 0$ and $q(t,r_+(t))\neq 0$, according to 
Theorem I, $\hat r_+(t)$ is a past outer spacelike horizon in the region 
$t\in[t_i,\, 0.44]$. For later times, one has ${\cal L}_+\Theta_- >0$ ($\hat 
r_+(t)$ changes character from outer to inner) and so it becomes a past inner 
timelike (Lorentzian) horizon in that region (i.e., for $t\in[t_1\simeq0.44,\, 
t_c\simeq 38.73]$), as predicted from the mentioned theorem. 
For times later than $t_c\simeq 38.73$ the NEC is not fulfilled in $\hat r_+ 
(t)$ and Theorem I does not apply. But, we verify that it changes back to 
spacelike character close to $t=t_f$. More precisely, it is an outer spacelike 
horizon in the interval $t\in [t_c,\, t_f].$

\item Along $\hat r_ -(t)$: The energy conditions are satisfied just very
close to the initial $t_i$, and are all violated for later times 
($t>t_2\simeq 0.35$ with the data of Fig.~\ref{fig:diagrama-Sinh23}).
Given that  $\Theta_+ > 0$, ${\cal L}_+\Theta_-<0$, and $q\neq 0$, 
it follows that $\hat r_-(t)$ is a past outer spacelike horizon in the 
region $t\in[0.31,\, 0.35]$.
For later times the NEC is violated and so Theorem I does not apply.
However, it can be shown that $\hat r_-(t)$ is a past outer timelike horizon
for intermediate times, in the interval $t\in [0.35,\,58.99]$, changing back
to inner spacelike character close to $t_f$, 
in the interval $t\in[59.0,\,62.82] $.

\end{enumerate}

\subsection{The case of Sec. \texorpdfstring{\ref{sec:dust}}{Lg} }

In the case of dust matter, the ratio  $n\left(t,r_+(t)\right)$
[see Eq.~\eqref{eq:ratio}], is larger
than unity for all times after the horizon formation along $r=\hat r_+(t)$,
while $n(t,\hat r_-(t))$ is larger than unity just
for very early times after horizons formation. In the situation of 
Fig.~\ref{fig:diagrama-dust}, the horizons are formed at $t=t_i
=4096\,m^3/27\, t_0^2\simeq 151.70$,
with values $\hat r_\pm(t_i)=4.0$ and last forever.
In fact, $\hat r_+(t)\to 3t_0\left(t/t_0\right)^{1/3}/2$ and 
$\hat r_-(t)$ tends to $2m$ as $t\to \infty$.

\begin{enumerate}[(1)]
 \item Along $\hat r_+(t)$:  The NEC, the WEC and the SEC are satisfied all 
along $\hat r_+ (t) $. The Lie derivative ${\cal L}_+\Theta_-\big|_{\hat 
r_+(t)}$ changes sign at $t_c= 216.0$. Close to $t_i$, more precisely, in the 
interval $[t_i,\, t_c]$,  the condition ${\cal L}_+\Theta_-<0$ holds (besides 
$\Theta_+ > 0$ and $q\neq 0$) and then $\hat r_+(t)$ is a past outer spacelike 
horizon in that region. For later times, ${\cal L}_+\Theta_- >0$ and so, in 
the interval $t\in[t_c,\, \infty) $, the $\hat r_+(t)$ branch changes 
character to a past inner timelike horizon, as predicted from Theorem I.   
 
\item Along $\hat r_ -(t)$: The energy conditions are satisfied
just close to the initial ($t_i)$, and are all violated for late times 
($t>t_m\simeq 172.47$ with the data of Fig.~\ref{fig:diagrama-Sinh23}). Given 
that  $\Theta_+ > 0$, ${\cal L}_+\Theta_-<0$, and $q\neq 0$, $\hat r_-(t)$ is 
a past outer spacelike horizons in the region $t\in [t_i,\, t_m]$. For times 
larger than $t_m$, the NEC is violated and Theorem I does not apply, but it 
can be shown that $\hat r_ -(t)$ is a past outer timelike horizon for all 
times after $t_c$.

\end{enumerate}

\subsection{The case of Sec. \texorpdfstring{\ref{sec:stiff}}{Lg}}

In the case of stiff matter, the ratio  $n\left(t,r_+(t)\right)$,
along $r=\hat r_+(t)$, is larger than unity
for all times after the horizon formation, 
while $n(t,\hat r_-(t))$ is larger than unity just at
early times after horizons formation. In the situation of 
Fig.~\ref{fig:diagrama-stiff-matter}, the horizons are formed a
t $t=t_i\simeq 4.35$,
with values $\hat r_\pm(t_i)=4.0$ and last forever. 
In the limit of very large times $\hat r_+(t)$ goes as 
$3t_0\left(t/t_0\right)^{2/3}/2$ while $r_-(t)\to 2m$.

\begin{enumerate}[(1)]
 \item Along $\hat r_+(t)$:   
The NEC and the WEC are satisfied for all times on $\hat r_+(t)$ since it is
formed, but we can show that the SEC is not satisfied along it.
Once the NEC is satisfied and, moreover, 
$\Theta_+ > 0$, ${\cal L}_+\Theta_-<0$, and $q\neq 0$ all along $\hat r_+(t)$, 
according to Theorem I,  it is a past outer spacelike horizon 
(see Fig.~\ref{fig:diagrama-stiff-matter}).

\item Along $\hat r_ -(t)$:  The NEC and the WEC are satisfied at
the beginning ($t\gtrsim t_i$), but all energy conditions are violated
for times larger that $t_c\simeq 6.70$. 
As in the case of the last section, at times close to $t_i$ one has 
${\cal L}_+\Theta_- <0$ and $q\neq 0 $, and since the NEC is satisfied along
$\hat r_-(t)$, Theorem I implies it is a past outer spacelike horizon
in the region $t\in[4,\, 6.70]$. For later times, the NEC is violated 
and so the hypothesis of the above mentioned theorem are not
fulfilled, but it is verified that $\hat r_-(t)$  is a past outer 
timelike horizon in that region.

\end{enumerate}

\subsection{The case of Sec. \texorpdfstring{\ref{sec:linear}}{Lg}}

In this case, $a(t) = t/t_0$, if $t_0\geq 8m$
the trapping horizons are formed at $t=0 $ and persist for all times
with constant values.
In the particular case of Fig.~\ref{fig:diagrama-linear}, 
where $t_0=10m$, $m=1$, the values are $\hat r_+(t)\simeq 7.24$
and $\hat r_-(t)\simeq 2.76$.

\begin{enumerate}[(1)]
 \item Along $\hat r_+(t)$:   
The NEC is satisfied for all times, and, moreover, considering that
$\Theta_+ > 0$, ${\cal L}_+\Theta_->0$, and $q\neq 0$ at all times on 
$\hat r_+(t)$. Thus, according to the relevant theorem one has that it is
a past inner timelike horizon (see Fig.~\ref{fig:diagrama-linear}).

\item Along $\hat r_ -(t)$:  The NEC is violated at all times along
this branch of the trapping horizon and the predictions of Theorem I
cannot be applied. However, since one has $\Theta_+=0$ and
${\cal L}_+\Theta_- <0$ it results that $\hat r_ -(t)$ is a past outer
horizon, and it can also be shown that it is everywhere timelike.

\end{enumerate}

\subsection{The case of Sec. \texorpdfstring{\ref{sec:tanh23}}{Lg} }

In the situation of Fig.~\ref{fig:diagrama-tanh23}, the horizons 
$\hat r_\pm(t)$ are formed at $t\simeq 1.498$ with values 
$\hat r_\pm= 4.0$, and last forever. Their asymptotic values for large times
 are $\hat r_-(t) = 2m$ and $\hat r_+(t) = 3 t_0 e^{2t/t_0}/2$.
For such a case,
we have shown that $n(t,\hat r_+\left(t)\right)\geq 1$, and 
that the NEC, the WEC and the SEC are obeyed all along $r_+(t)$. 
The same holds along $\hat r_-(t)$.

\begin{enumerate}[(1)]
 \item Along $\hat r_+(t)$:   
The NEC is satisfied for all times, and, moreover, considering that
$\Theta_+ > 0$, ${\cal L}_+\Theta_-<0$, and $q\geq 0$ at all times on
$\hat r_+(t)$, according to Theorem I it is a past outer 
spacelike horizon.

\item Along $\hat r_ -(t)$:  The NEC holds at all times along
this branch of the trapping horizon and the predictions of the mentioned 
theorem, together with the conditions 
${\cal L}_+\Theta_- <0$ and $q\neq0$ on $\hat r_-(t)$, imply it is a past
outer spacelike horizon (see Fig.~\ref{fig:diagrama-tanh23}).

\end{enumerate}

\subsection{The cases of Sec. \texorpdfstring{\ref{sec:tanh}}{Lg} }

For $a(t)=\tanh\left(t/t_0\right)$ it is convenient to split the analysis
into three different classes
according to the values of $t_0$: ($a$) $t_0<4m$; ($b$) $4m< t_0 < 8m$; and 
($c$) $t_0>8m$.

\subsubsection{ The case \texorpdfstring{$t_0< 4m$}{Lg}  }

In the situation of Fig.~\ref{fig:diagrama-Tanh3}, with $t_0=3m$, $m=1$,
the horizons are formed at $t_i\simeq 3.22$ with values $\hat r_\pm(t_i)
=4.0$ and last forever. In the limit $t\to\infty$, $\hat r_+(t)$ tends to 
$t_0\ e^{2t/t_0}$
while $\hat r_-(t)$ tends to $2m$.

\begin{enumerate}[(1)]
 \item Along $\hat r_+(t)$:   
The NEC is satisfied for all times, and, moreover, considering that
$q\neq 0$ and ${\cal L}_+\Theta_-<0$ at all times on $\hat r_+(t)$,
according to the relevant theorem it is a past outer 
spacelike horizon.

\item Along $\hat r_ -(t)$:  The NEC holds at all times along
this branch of the trapping horizon and the predictions Theorem I,
together with the conditions $q\neq 0$ and 
${\cal L}_+\Theta_- <0$ on $\hat r_-(t)$, which are fulfilled at all times on 
$\hat r_-(t)$ imply it is a past outer spacelike
horizon. In the limit $t\to\infty$ it tends to the lightlike horizon
$r=2m$ (see Fig.~\ref{fig:diagrama-Tanh3}).
\end{enumerate}

\subsubsection{ The case \texorpdfstring{$4m< t_0< 8m$}{Lg}  }

In the situation of Fig.~\ref{fig:diagrama-Tanh6}, with $t_0=6m$, $m=1$,
the horizons are formed at $t_i\simeq 3.30$ with values $\hat r_\pm(t_i)
=4.0$ and last forever. In the limit $t\to\infty$, $\hat r_+(t)$ tends to
$t_0\, e^{2t/t_0}$,
while $\hat r_-(t)$ tends to $2m$.

\begin{enumerate}[(1)]
 \item Along $\hat r_+(t)$:   
The NEC is satisfied for all times and so Theorem I applies.
Considering that $q\neq 0$  and ${\cal L}_+\Theta_-<0$ 
at all times on $\hat r_+(t)$ the result is a past outer 
spacelike horizon.

\item Along $\hat r_ -(t)$:  The NEC holds at initial times after horizons
formation, from $t_i\simeq 3.30$ to $t_c\simeq 3.95$ along
this branch of the trapping horizon. Moreover, one has the conditions $q\neq 0$ and
${\cal L}_+\Theta_- >0$ on $\hat r_-(t)$, which implies that it is a past
outer spacelike horizon (see Fig.~\ref{fig:diagrama-Tanh6})
 in that region. Later than $t\simeq 3.95$ the NEC is violated
and the horizon changes character to timelike, being a past outer horizon in the
interval $t\in[t_c,\,\infty)$. In the limit $t\to\infty$ 
it asymptotes the lightlike horizon $r=2m$, a region where the NEC 
holds also on $\hat r_-(t)$. 
\end{enumerate}

\subsubsection{ The case \texorpdfstring{$t_0> 8m$}{Lg}  }

In the situation of Fig.~\ref{fig:diagrama-Tanh10}, with $t_0=10m$, $m=1$,
the horizons are formed at $t_i=0.0$ with values $\hat r_+(t_i) \simeq 7.2$
and $r_-(t_i)\simeq 2.8$ and last forever. 
In the limit $t\to\infty$, $\hat r_+(t)$ tends to $t_0\, e^{2t/t_0}$,
while $\hat r_-(t)$ tends to $2m$.

\begin{enumerate}[(1)]
 \item Along $\hat r_+(t)$:   
The NEC is satisfied for all times and Theorem I can be applied. The Lie 
derivative ${\cal L}_+\Theta_-$ is positive at initial times but changes to 
negative values at $t=t_c\simeq 5.93$. Then, $\hat r_+(t)$ is a past outer 
timelike horizon in the interval $t\in [0,\, t_c]$, and is a past inner 
spacelike horizon in the region $t\in [t_c,\, \infty)$.

\item Along $\hat r_ -(t)$:  The energy conditions are violated. In particular 
the NEC does not hold along this branch of the trapping horizon and the 
predictions of the mentioned theorem do not apply here. However, it is 
verified that $q\neq 0$ and ${\cal L}_+\Theta_- <0$ all along $\hat r_-(t)$ 
and so it is a past outer timelike horizon.

\end{enumerate}

\subsection{The cases of Sec. \texorpdfstring{\ref{sec:truncated}}{Lg} }

As commented at the end of Sec.\ref{sec:truncated}, these are approximate 
models that have been implemented to simulate a phase change on the accreting 
process onto the central object. As a first approximation, we truncated the 
scale factor in such a way its rate $\dot a(t)$ vanishes at very late times. 
With such a choice, the character of the trapping horizons remains majorly the 
same as the original models. Moreover, even in these crudely approximated 
models we used to draw the causal diagrams, the horizon functions $\hat 
r_\pm(t)$ have no jumps with time. The modifications happen just in the limit 
$t\to\infty$. For instance, in the model of Eq.~\eqref{eq:power2cutoff} the 
only change is in the final value of $\hat r_+(t)$ which tends to infinity in 
the original model, while it approaches a constant (though with a very large 
value) in the modified model. The energy conditions on it and the 
chronological character, compatible with Theorem I, do not change. Also, the 
character of the branch $\hat r_-(t)$ remains the same.

\end{document}